\documentclass[12pt,pdfstartview=FitH]{article}

\usepackage{graphicx} % Required for inserting images
\usepackage{amssymb}
\usepackage{amsmath}
\usepackage{jheppub}
\usepackage{bbm}
\usepackage{bm} % Bold math symbols. Remove if not rendering properly.
\usepackage{enumitem}
\usepackage{physics}
\usepackage{xcolor}
\usepackage{empheq}
\usepackage{tikz}
\usepackage{picture}
\usepackage[Smaller,makeroom,thicklines]{cancel}
 
\usepackage{Shortcuts}

\title{Proper time to singularity and thermal correlators}
\author{Kaustubh Singhi}
\affiliation{International Centre for Theoretical Sciences (ICTS-TIFR), \\ Tata Institute of Fundamental Research, \\ Shivakote, Hesaraghatta, Bangalore 560089, India.}
\emailAdd{kaustubh.singhi@icts.res.in}

\date{\today}

\abstract{We study certain higher point thermal correlators of heavy and light scalar primaries in a holographic CFT. Assuming simple self-interactions and couplings of the scalars in the bulk theory, we show that the thermal correlators contain a signature of the proper time to singularity from the horizon. The key idea is to use WKB approximations for the propagators and evaluate the bulk integrals using saddle point method.}

\begin{document}
\maketitle

\section{Introduction}
Given the success of the AdS/CFT correspondence in correctly computing CFT correlation functions from entirely bulk supergravity calculations, it is natural to ask which features of the AdS bulk are present in the boundary CFT. In particular, the semiclassical bulk theory permits various kinds of black hole solutions, and so one can ask whether (and how) the geometrical features of black holes are encoded in the dual CFT. \\

A reason that this question is not straightforward to answer is that the statement of the duality is best stated in the Euclidean theories. In presence of a black hole, the bulk Euclidean theory lives only on the exterior of the black hole. In this case, the correspondence relates Euclidean CFT correlators to those of fields living in the exterior geometry. \\

Regardless of the fact that the Euclidean bulk only consists of exterior geometry, the metric is still obtained from continuation of the Lorentzian one. So, it is often expected that geometrical features such as the black hole interior and singularity, present in the full Lorentzian geometry, have to be obtained by some analytic continuation of the Euclidean CFT correlators. This connection has been explored in the early days of AdS/CFT \cite{Kraus:2002iv,Fidkowski:2003nf,Festuccia:2005pi} and recently in \cite{Horowitz:2023ury,Ceplak:2024bja}. The key result in these papers is a signature of the black hole singularity in the analytically continued Euclidean correlators. \\

In contrast, Grinberg and Maldacena \cite{Grinberg:2020fdj} have argued that in the presence of a higher derivative coupling of a heavy scalar, such as  Weyl tensor squared, thermal one point function of the dual CFT primary contains a signature of the proper time to singularity from the horizon. The bulk calculation is performed in the large mass (WKB) limit, where the scalar wave equation can be solved using geodesics \cite{Louko_2000}. In this limit, some bulk integrals can be performed using saddle point approximation, and so the calculation reduces to finding the location of saddle points in the bulk geometry. \\

Grinberg and Maldacena's proposal has renewed interest in thermal correlators in holographic CFTs and signatures of the black hole singularity. David and Kumar \cite{David:2022nfn} have strengthened their claim by performing exact calculations in different black hole backgrounds while also considering a Gauss-Bonnet coupling of the scalar instead of Weyl squared one. They have also tried addressing the problem from the CFT side \cite{David:2022nfn,David:2023uya}. To probe richer physics than that available from the one point function, there have been several attempts to generalize the claim to higher point functions \cite{Rodriguez-Gomez:2021pfh,Rodriguez-Gomez:2021mkk,Krishna:2021fus,Georgiou:2022ekc,Georgiou:2023xpg}. These attempts focus on using geodesics to study the leading structure of thermal two and three point functions of CFT primaries of large conformal dimension. \\

In this paper, we look for geometrical features of the bulk in more general higher point thermal correlators of CFT primaries of large and small conformal dimensions. To this end, we introduce heavy and light bulk scalar fields in a black brane background and consider bulk interactions that will source higher point functions of the dual operators. We argue that to obtain the information of proper time to singularity in thermal correlators, it is not necessary to couple the heavy scalar to higher derivative curvature terms --- even simple self-interaction terms of the heavy scalar, or interactions of the heavy and light scalar are sufficient to give this signature of singularity. Our emphasis in this paper is on keeping the first two terms in the WKB expansion and then directly performing the bulk integrals using saddle point method. This is in contrast to earlier attempts where the focus has been to replace the calculation with some geodesic computations. \\

The organization of the paper is as follows. In sections \ref{Section: setup} and \ref{Section: preliminaries}, we provide the requisite background and obtain the WKB expressions for the scalar propagators required to compute thermal correlators. To my knowledge, with the exception of an appendix in \cite{Grinberg:2020fdj}, previous computations have featured only the leading order WKB solution, i.e. the geodesic approximation, for the boundary-bulk propagators. Since we require both boundary-bulk and bulk-bulk propagators at next-to-leading order in WKB, we spend some time in deriving these.\footnote{See \cite{Krishna:2021fus} for a somewhat different treatment of the bulk-bulk propagator.} In sections \ref{Section: n point function} and \ref{Section: heavy light correlator}, we calculate appropriate higher point thermal correlators of CFT primaries and show that these too contain the information of proper radial time to singularity. Contributions to thermal correlators from Witten diagrams containing bulk-bulk propagators are considered in section \ref{Section: bulk propagator contributions} along with arguments for the appearance of the time to singularity in these contributions as well. {\color{black} In section \ref{Section: other symmetric solutions}, we extend the proposal to other black hole backgrounds and show that in these cases as well higher point thermal correlators probe the interior geometry and singularity.} Finally, we conclude with some discussions on our results.

% ---------------------------------------------------------------------

\section{Setup}
\label{Section: setup}
Consider a $d+1$-dimensional black brane AdS geometry with heavy and light probe scalar fields, $\phi$ and $\chi$, respectively. These scalars are dual to conformal primaries $\ophi$ and $\ochi$ in a thermal state of the holographic CFT. Using the AdS/CFT dictionary (refer \cite{Freedman_Van_Proeyen_2012} for details on AdS/CFT in the supergravity limit), the relation between the conformal dimensions of $\ophi$ and $\ochi$ to the masses, $\mphi$ and $\mchi$, of bulk scalars $\phi$ and $\chi$, respectively, is given by\footnote{The AdS length scale has been set to unity.}
\begin{align}
    \label{Conformal dimension 1}
    \Dphi & = \frac{d}{2} + \sqrt{\frac{d^{2}}{4} + \mphi^{2}}, \\
    \label{Conformal dimension 2}
    \Dchi & = \frac{d}{2} + \sqrt{\frac{d^{2}}{4} + \mchi^{2}}.
\end{align}
By taking $\mphi$ large, we are studying correlators of CFT primaries of large conformal dimension. \\

In \cite{Grinberg:2020fdj}, Grinberg and Maldacena considered a Weyl squared coupling of the heavy scalar in the bulk (as in \eqref{Action 5}). From the point of view of AdS/CFT, the Weyl squared coupling sources the thermal one point function of the dual primary operator. In the probe limit, it is simply given by a bulk integral of the boundary-bulk $\phi$ propagator weighted by the value of the Weyl squared term (see Figure \ref{Witten diagram 5}). Using WKB and saddle point methods, it can be argued that the thermal one point function of $\ophi$ takes the following form
\begin{gather}
    \label{One point function 1}
    \ev{\ophi}_{\beta} \sim \exp[-\mphi(\ell_{h} + i\tau_{s})].
\end{gather}
In this expression, $\ell_{h}$ is the regularized distance between the black hole horizon and the boundary, and $\tau_{s}$ is the proper time a radially infalling geodesic observer takes to reach the singularity from the horizon. The above form has been verified by exact computation as well. Calculation of the thermal one point function has been reviewed in appendix \ref{Section: one point function}. \\

A form similar to \eqref{One point function 1} has been conjectured for a variety of symmetric black hole solutions. In the case of black holes with an inner horizon, instead of $i\tau_{s}$, the one point function features the proper radial time between the two horizons ($i\tau_{in}$) and the proper radial distance between the inner horizon and the singularity ($\ell_{sing}$). David and Kumar \cite{David:2022nfn} have verified by exact computation this conjecture in the case of a charged black hole in a particular large $d$ (spacetime dimension) limit. They have also obtained similar results for the thermal one point function by considering a Gauss-Bonnet coupling instead of Weyl squared. \\

In this paper, we establish that to see a geometrical feature such as the time to singularity from the thermal CFT, it is \emph{not} necessary to introduce a bulk coupling of the scalar with higher derivative curvature terms. In fact, even correlators sourced by simple self-interaction terms or scalar exchange couplings contain the signature of time to singularity. \\

We begin by considering two classes of bulk couplings. Firstly, we take a self-interaction term of the form $(\phi)^{n}$ for $n \geq 3$. In the dual CFT, the $n$-point function of the primary operator $\ophi$ gets a tree level contribution from this term (see Figure \ref{Witten diagram 1}). Secondly, we consider a coupling between the heavy and light scalars. We take a `heavy-light' interaction of the kind $(\phi)^{N}(\chi)^{M}$, for positive integers $N,M$. Such a coupling sources correlators of the schematic kind $(\ophi)^{N}(\ochi)^{M}$ (see Figure \ref{Witten diagram 2}). \\

The Euclidean bulk action is written as
\begin{gather}
    \label{Action 1}
    S[\phi,\chi] = S_{\rm free}[\phi] + S_{\rm free}[\chi] + S_{\rm int}[\phi,\chi].
\end{gather}
The free part is simply the Klein-Gordon action
\begin{align}
    \label{Action 2}
    S_{\rm free}[\phi] & = \frac{1}{2}\int{d^{d+1}x}\sqrt{g}\l[\p_{\mu}\phi\p^{\mu}\phi + \mphi^{2}\phi^{2} \r], \\
    \label{Action 3}
    S_{\rm free}[\chi] & = \frac{1}{2}\int{d^{d+1}x}\sqrt{g}\l[\p_{\mu}\chi\p^{\mu}\chi + \mchi^{2}\chi^{2} \r],
\end{align}
and the interaction, as just discussed, is chosen to be
\begin{gather}
    \label{Action 4}
    S_{\rm int}[\phi,\chi] = \int{d^{d+1}x}\sqrt{g}\l[\frac{\lambda_{n}}{n!}(\phi)^{n} + \frac{\kappa_{N,M}}{N!M!}(\phi)^{N}(\chi)^{M}\r].
\end{gather}
To avoid cumbersome expressions, factors of $16\pi G$ have been suppressed. Setting the AdS length scale to 1, the bulk Euclidean metric for the black brane is given by 
\begin{gather}
    \label{Black brane metric 1}
    ds^{2} = \frac{1}{z^{2}}\l(f(z)dt^{2} + \frac{1}{f(z)}dz^{2} + d\vec{x}^{2} \r) \quad ; \quad f(z) = 1 - \frac{z^{d}}{z_{h}^{d}}.
\end{gather}
The Euclidean bulk geometry consists only of the exterior of the black hole, i.e. of the region $0 < z \leq z_{h}$, and all bulk integrals run over this range only. \\

Thermal correlators in the CFT can be evaluated holographically by computing integrals over bulk propagators. For example, the leading contribution from the self-interaction in \eqref{Action 4} to the thermal $n$-point function of $\ophi$ is given by the following bulk integral
\begin{gather}
    \label{n point function 1}
    \ev{\ophi(k_{1}) \cdots \ophi(k_{n})}_{\beta} = \lambda_{n}\int{dz}\sqrt{g(z)}G_{k_{1}}^{\phi}(z) \ldots G_{k_{n}}^{\phi}(z).
\end{gather}
Similarly, the heavy-light interaction sources the following thermal correlator
\begin{align}
    \begin{split}
    \label{Heavy light correlator 1}
    \ev{\ophi(k_{1}) \cdots \ophi(k_{N})\ochi(q_{1}) \cdots \ochi(q_{M})}_{\beta} & \\
    & \hspace{-3.75cm} = \kappa_{N,M}\int{dz}\sqrt{g(z)}G_{k_{1}}^{\phi}(z) \ldots G_{k_{N}}^{\phi}(z)G_{q_{1}}^{\chi}(z) \ldots G_{q_{M}}^{\chi}(z).
    \end{split}
\end{align}
In the above expressions, the momentum space boundary-bulk propagators for the fields $\phi$ and $\chi$ are denoted, respectively, by $G_{k}^{\phi}(z)$ and $G_{q}^{\chi}(z)$.\footnote{Since the black brane background has translation isometries in all non-radial directions, momentum space fields and correlators are related to position space ones by a simple Fourier transform.} A factor of $(2\pi)^{d}$ and an overall momentum conserving delta function have been suppressed.

% ---------------------------------------------------------------------

\section{Preliminaries}
\label{Section: preliminaries}
In this section, we obtain the approximate expressions for the boundary-bulk and bulk-bulk propagators of the $\phi$ field in the large mass (WKB) limit. These WKB expressions will then be used to compute integrals of the kind \eqref{n point function 1} and \eqref{Heavy light correlator 1}. The bulk-bulk propagators will be of use for the computations in section \ref{Section: bulk propagator contributions}. Explicit expressions for $\chi$-propagators won't be required in the saddle point computations we perform.  \\

We introduce momentum $k \equiv (\o,\vec{k})$ corresponding to the translational isometries of the black brane backgrounds. Then, only the radial part of the wave equation remains to be solved.\footnote{A similar statement can be made for other spherically symmetric black hole backgrounds.} Now, there is a choice to be made on which our approximate WKB solutions \emph{will} depend: whether to scale the momentum with the mass or not. \\

The limit in which we work throughout this paper is where we take the mass $\mphi$ to be large while keeping the momentum fixed. In this case, the `leading' order WKB solutions do not get any momentum dependence. In particular, as we will see below, the turning point is still at the horizon in this limit. Since we compute only leading order exponential dependence on $\mphi$ of the correlators in question, it suffices to keep only the next-to-leading order WKB solutions. \\

The other limit, in which the momentum is scaled with a factor of $\mphi$ and then the large mass limit is taken, relates the WKB solutions to bulk geodesics. In this case, the turning point lies in the exterior region, so the saddle arguments to approximate the bulk integrals have to be made much more carefully. Even after restricting to a specific dimension and momentum values, it is not at all clear that the CFT correlators in this limit will give results similar to the ones obtained in the other limit. An analysis is performed in \cite{Rodriguez-Gomez:2021mkk} to compute thermal three point function by reducing the WKB solution to geodesics carrying generic momentum. \\

We emphasize that the relation between the WKB ansatz, like the one obtained in \eqref{WKB solution 1}, and the geodesic distances is only for the first term. The successive terms in the WKB expansion have to be obtained by solving the wave equation. Furthermore, if the turning point lies in the domain of the bulk integrals, as it does in the second limit, then a prescription has to be adopted for fixing the WKB solutions. This limit will be studied at a later time. Of course, it is perfectly reasonable to just study low modes in the large mass limit. Hence, we proceed with the first case in which the momentum is left untouched while taking the large mass limit. \\

The opposite limit --- taking the mass small (vanishing) and momentum large --- was studied in \cite{Amado:2008hw} in a different context. It will be interesting to see whether such a limit can also give rise to a signature of the singularity in the boundary correlators.

\subsection{WKB approximations}
We want to get approximate solutions to the Klein-Gordon equation for the heavy scalar, $\phi$, in the black brane background:
\begin{gather}
    \label{KG equation 1}
    \frac{1}{\sqrt{g}}\p_{\mu}(\sqrt{g}\p^{\mu}\phi) - \mphi^{2}\phi = 0,
\end{gather}
where $g_{\mu\nu}$ is the Euclidean metric \eqref{Black brane metric 1}. After going to Fourier space, the radial equation reads as
\begin{gather}
    \label{KG equation 2}
    z^{d+1}\p_{z}(z^{-d+1}f(z)\p_{z}\phi_{k}(z)) - \l(\frac{\o^{2}}{f(z)}z^{2} + \abs*{\vec{k}}^{2}z^{2} + \mphi^{2} \r)\phi_{k}(z) = 0
\end{gather}
with $k \equiv (\o,\vec{k})$. \\

Now, we find WKB solutions to \eqref{KG equation 2} in the large mass limit by taking the ansatz $\phi_{k}(z) \propto \exp(\mphi S(z))$, where $S(z)$ admits an expansion in $\frac{1}{\mphi}$. Since we are \emph{not} scaling the momentum by $\mphi$, the leading order WKB solution reads the same as in the case of vanishing momentum.\footnote{From hereon, by leading order WKB solution we mean the first two terms in the $\frac{1}{\mphi}$ expansion of $S(z)$.} The momentum contributes to $S(z)$ at order $\displaystyle\frac{1}{\mphi^{2}}$. \\

Substituting the WKB ansatz into the equation, we find the solutions for $S(z)$. They are conveniently written in the variable $w = \frac{z^{d}}{z_{h}^{d}}$
\begin{gather}
    \label{WKB solution 1}
    S(z) \equiv S(w) = \pm\ell_{w} + \frac{1}{4\mphi}\log(\frac{w^{2}}{1 - w}) + O\l(\frac{1}{\mphi^{2}} \r),
\end{gather}
where the $\pm$ sign corresponds to normalizable and non-normalizable solutions, respectively. $\ell_{w}$ is the regularized radial geodesic distance from the boundary to the bulk point $w$. It is evaluated by the integral
\begin{gather}
    \label{Regularized length 1}
    \ell_{w} = \frac{1}{d}\lim_{w_{c} \to 0}\l(\int_{w_{c}}^{w}{\frac{dw}{w\sqrt{1 - w}}} + \log w_{c}\r) = \ell_{h} - \frac{2}{d}\tanh^{-1}\sqrt{1 - w}.
\end{gather}
The identification $\ell_{h} \equiv \frac{2}{d}\log2$ has been made for the regularized distance to the horizon. \\

Some comments are in order concerning the validity of WKB approximation for the solutions. The second term in \eqref{WKB solution 1} gives the prefactor for the exponent. Because of fractional powers of $w$ and $1 - w$, it appears as if both the solutions have branch points at \emph{both} $w = 0,1$. But this is not quite correct, since the expansion itself will break down at the turning point, which in this case is at the horizon, $w = 1$. This can be seen more explicitly by solving the Klein-Gordon equation \eqref{KG equation 2} exactly for $\o,\vec{k} = 0$. The normalizable and non-normalizable solutions are respectively given by
\begin{align}
    \label{Normalizable solution 1}
    \phi_{0}^{\rm norm}(w) & = w^{\hphi}\hf(\hphi,\hphi;2\hphi;w), \\
    \label{Nonnormalizable solution 1}
    \phi_{0}^{\rm nn}(w) & = w^{\hphi}\hf(\hphi,\hphi;1;1 - w),
\end{align}
where we have introduced the parameter $\displaystyle \hphi \equiv \frac{\Dphi}{d}$. \\

It is clear from \eqref{Normalizable solution 1} that the normalizable solution has a branch point at $w = 0$ because of the $w^{h}$ term and a branch point at $w = 1$ due to the hypergeometric. In contrast, from \eqref{Nonnormalizable solution 1} it can be seen that the non-normalizable solution only has a branch point at $w = 0$. This means that in the WKB solutions, \emph{part} of the branch point at $w = 1$ is indeed a spurious artefact of the solution being invalid near the turning point. \\

It may be possible to account for this spurious feature by performing an `exact WKB' computation \cite{kawai2005algebraic}, or even by solving the equation near the turning point and matching it against the WKB solution. But for the calculations in this paper, it is not necessary to do so. In the saddle point analysis that follows, we will simply restrict to using the WKB solutions in the region away from the turning point so that the approximation still holds. We hence only consider saddle points away from the turning point. \\

The large mass expression for the solutions at vanishing momentum can also be got by using an integral representation for the hypergeometric and then evaluating the integral in saddle point method. As a consistency check, we give an alternative derivation of the WKB expression of the boundary-bulk propagator (proportional to the non-normalizable solution) in appendix \ref{Section: consistency check 1}. \\

Now that we have the solutions, we need expressions for the boundary-bulk and bulk-bulk propagators. Since we are only interested in correctly producing the exponent in the saddle point approximation, we can suppress some overall factors and numerical constants. \\

The boundary-bulk propagator is simply proportional to the non-normalizable solution. In the WKB limit taken to obtain \eqref{WKB solution 1}, the boundary-bulk propagator is simply given as
\begin{gather}
    \label{Boundary-bulk propagator 1}
    G_{k}^{\phi}(w) \sim \sqrt{\frac{w}{\sqrt{1 - w}}}e^{-\mphi\ell_{w}}\l(1 + O\l(\frac{1}{\mphi} \r) \r),
\end{gather}
where $\ell_{w}$ is given in \eqref{Regularized length 1}. The normalization of the boundary-bulk propagator is obtained by the factor of the horizon distance $\ell_{h}$. \\

Given the two independent solutions in \eqref{WKB solution 1}, the bulk-bulk propagator can be constructed in the usual way using the Heaviside function
\begin{gather}
    \label{Bulk-bulk propagator 1}
    \mcg_{k}^{\phi}(w_{1};w_{2}) \sim \sqrt{\frac{w_{1}w_{2}}{\sqrt{1 - w_{1}}\sqrt{1 - w_{2}}}}\l[e^{-\mphi\ell_{w_{1}}}e^{\mphi\ell_{w_{2}}}\Theta(w_{1} > w_{2}) + e^{-\mphi\ell_{w_{2}}}e^{\mphi\ell_{w_{1}}}\Theta(w_{1} < w_{2})\r].
\end{gather}
Since $\ell_{w}$ is a monotonically increasing function of $w$ as we move from the boundary to the horizon, the bulk-bulk propagator can be written more succinctly as
\begin{gather}
    \label{Bulk-bulk propagator 2}
    \mcg_{k}^{\phi}(w_{1};w_{2}) \sim \sqrt{\frac{w_{1}w_{2}}{\sqrt{1 - w_{1}}\sqrt{1 - w_{2}}}}e^{-\mphi\abs{\ell_{w_{1}} - \ell_{w_{2}}}}\l(1 + O\l(\frac{1}{\mphi} \r) \r).
\end{gather}

The prefactors in \eqref{Boundary-bulk propagator 1}-\eqref{Bulk-bulk propagator 2} coming from the leading order WKB solution are often ignored. A reason is that these prefactors `shift' the saddle points in a way that only subleading contributions get affected. But we keep these factors along to fix the saddle locations. This will become clear when we explicitly evaluate the integrals for the correlation functions using saddle point method.

% ---------------------------------------------------------------------

\section{$n$-point function}
\label{Section: n point function}
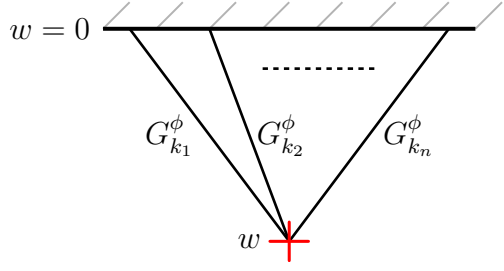
\begin{figure}[!t]
    \centering
    % Witten diagram for n+1 point correlator
    \begin{tikzpicture}[x=1pt,y=1pt,yscale=-1,xscale=1]
    % \path (0,451); %set diagram left start at 0, and has height of 451
    
    % Base boundary line
    \draw [line width = 1.5] (100,110) -- (240,110);
    \draw (80,110) node {$w = 0$};
    
    % Low opacity boundary lines
    \draw [opacity = 0.3, line width = 0.75] (100,110) -- (110,100);
    \draw [opacity = 0.3, line width = 0.75] (120,110) -- (130,100);
    \draw [opacity = 0.3, line width = 0.75] (140,110) -- (150,100);
    \draw [opacity = 0.3, line width = 0.75] (160,110) -- (170,100);
    \draw [opacity = 0.3, line width = 0.75] (180,110) -- (190,100);
    \draw [opacity = 0.3, line width = 0.75] (200,110) -- (210,100);
    \draw [opacity = 0.3, line width = 0.75] (220,110) -- (230,100);
    \draw [opacity = 0.3, line width = 0.75] (240,110) -- (250,100);
    
    % Heavy scalar boundary-bulk propagator
    \draw [line width = 1] (110,110) -- (170,190);
    \draw (125,150) node {$G_{k_{1}}^{\phi}$};
    \draw [line width = 1] (140,110) -- (170,190);
    \draw (167,150) node {$G_{k_{2}}^{\phi}$};
    \draw [dash pattern = on 2pt off 2pt, line width = 1] (160,125) -- (203,125);
    \draw [line width = 1] (230,110) -- (170,190);
    \draw (215,150) node {$G_{k_{n}}^{\phi}$};

    % Self interaction vertex
    \draw (170,190) node [font = \LARGE, color={rgb, 255:red, 255; green, 0; blue, 0}] {$\bm{+}$};
    \draw (155,190) node {$w$};
     
    \end{tikzpicture}

    \caption{Witten diagram for the $n$-point function of $\ophi$ being sourced holographically by a self-interaction term of the dual scalar. The interaction vertex has been marked and the bulk point is integrated over the exterior region. The $\phi$ propagators are denoted by solid lines. Momenta carried by the propagators are indicated and are small in the sense that they \emph{do not} scale by the mass of the heavy scalar.}
    
    \label{Witten diagram 1}
\end{figure}

We begin by computing the thermal $n$-point function of a CFT primary, $\ophi$, of large conformal dimension, $\Dphi$. Such a correlator is sourced holographically by an $n$-point bulk self-interaction of the dual heavy scalar, $\phi$, as introduced in the action \eqref{Action 4}. The Witten diagram for the thermal correlator is illustrated in Figure \ref{Witten diagram 1}, and the corresponding bulk integral over the Euclidean black brane geometry is given by \eqref{n point function 1}. Shifting to the coordinate $w = \frac{z^{d}}{z_{h}^{d}}$ and plugging in the measure, the integral reads
\begin{gather}
    \label{n point function 2}
    \ev{\ophi(k_{1}) \cdots \ophi(k_{n})}_{\beta} \sim \lambda_{n}\int_{0}^{1}{\frac{dw}{w^{2}}}G_{k_{1}}^{\phi}(w) \ldots G_{k_{n}}^{\phi}(w).
\end{gather}

Recall that we are looking for signatures of the black hole singularity in the higher point thermal correlators. In the large mass limit, the propagators get linked to geodesics through the WKB solutions. We are interested in extracting the exponent in the large mass limit of the resulting integrals. Hence, we will not keep track of overall factors. \\

Substituting the WKB expression \eqref{Boundary-bulk propagator 1} for the boundary bulk propagator, and bringing all integrand factors in the exponent, the bulk integral reads
\begin{gather}
    \label{n point function 3}
    \ev{\ophi(k_{1}) \cdots \ophi(k_{n})}_{\beta} \sim \lambda_{n}\int_{0}^{1}{dw}\exp[-n\mphi\ell_{w} + \frac{n}{4}\log(\frac{w^{2}}{1 - w}) - \log w^{2}].
\end{gather}

The leading behaviour of the integral in \eqref{n point function 3} can be computed easily in saddle point method. The saddle equation reads
\begin{gather}
    \label{n point saddle 1}
    -n\mphi\eval{\p_{w}\ell_{w}}_{w^{*}} + \frac{n}{4}\l(\frac{2}{w^{*}} + \frac{1}{1 - w^{*}}\r) - \frac{2}{w^{*}} = 0,
\end{gather}
which, after putting in \eqref{Regularized length 1}, simplifies to
\begin{gather}
    \label{n point saddle 2}
    -\frac{n\mphi}{d}\frac{1}{w^{*}\sqrt{1 - w^{*}}} + \frac{2n - 8 - (n - 8)w^{*}}{4w^{*}(1 - w^{*})} = 0.
\end{gather}
At large $\mphi$, the leading behaviour of the saddle location is easily obtained as\footnote{\label{footnote: spurious saddle} There is another solution near $w = 1$ of equation \eqref{n point saddle 2}, but as discussed earlier, the WKB approximation is not valid here. Hence, this saddle should be ignored to avoid spurious results.}
\begin{gather}
    \label{n point saddle 3}
    w^{*} \approx -\l(\frac{4n}{n - 8}\frac{\mphi}{d}\r)^{2}.
\end{gather}
Now, to obtain the leading exponential behaviour of the integral in \eqref{n point function 3}, we simply need to evaluate the integrand at the saddle point. Since the saddle \eqref{n point saddle 3} lies at infinity, we run into an ambiguity in defining the regularized length $\ell_{w}$ at this point. One way to resolve this ambiguity is by giving $\mphi$ a negative imaginary part, so that for large $\mphi$, the expression \eqref{Regularized length 1} yields
\begin{gather}
    \label{Regularized length 2}
    \ell_{w^{*}} \approx \ell_{h} + i\tau_{s},
\end{gather}
where we have made the identification $\tau_{s} = \frac{\pi}{d}$ for the radial proper time to singularity. \\

Thus, the $n$-point function evaluates to the following form
\begin{gather}
    \label{n point function 4}
    \ev{\ophi(k_{1}) \cdots \ophi(k_{n})}_{\beta} \sim \lambda_{n}\exp[-n\mphi(\ell_{h} + i\tau_{s})] \times \text{powers of } \mphi.
\end{gather}

\subsection{Steepest descent analysis for the $n$-point function}
\label{Section: SDA 1}
We provide a detailed steepest descent analysis of the integral in \eqref{n point function 3}, thus justifying the expression \eqref{n point function 4}. The analysis presented here generalizes the treatment of the one-point function in section 4.3 of \cite{Grinberg:2020fdj}. \\

To visualize the saddle(s) contributing to the $n$-point function and their respective steepest descent contours, it is convenient to change variables to
\begin{gather}
    \label{SDA 1}
    w = \sech^{2}\frac{\rho}{2}.
\end{gather}
Substituting this in \eqref{Regularized length 1} gives the geodesic distance as a function of $\rho$
\begin{gather}
    \label{SDA 2}
    \ell_{\rho} = \ell_{h} - \frac{\rho}{d}.
\end{gather}
Taking into account the Jacobian of \eqref{SDA 1}, the integral \eqref{n point function 3} for the $n$-point function can be written as
\begin{gather}
    \label{SDA 3}
    \ev{\ophi(k_{1}) \cdots \ophi(k_{n})}_{\beta} \sim \lambda_{n}\int_{\mcc}{d\rho}\exp[-n\mphi\ell_{\rho} + \l(1 - \frac{n}{2}\r)\log(\sinh\frac{\rho}{2}\cosh\frac{\rho}{2})].
\end{gather}
In writing \eqref{SDA 3}, we have not explicitly specified the integration contour $\mcc$. This is because the integral on the original real contour $\rho \in (0,\infty)$ diverges for large $\mphi$. This also happens to be the case for the one-point function computed in \cite{Grinberg:2020fdj}. The resolution is to modify the integration contour so that at large $\abs{\rho}$ the exponent is decaying. For this to happen, $\mcc$ must asymptote in a direction such that 
\begin{gather}
    \label{SDA 4}
    \abs{\arg(\mphi\rho)} > \frac{\pi}{2}.
\end{gather}

Now, we come to finding the saddles of \eqref{SDA 3}. In the large $\mphi$ limit, these are solutions to the equation (ignoring the near horizon spurious saddles; see footnote \ref{footnote: spurious saddle})
\begin{gather}
    \label{SDA 5}
    \tanh\frac{\rho}{2} \approx \frac{4n}{n - 2}\frac{\mphi}{d}.
\end{gather}
Inverting the equation gives
\begin{gather}
    \label{SDA 6}
    \rho = -(2k + 1)i\pi + \eta
\end{gather}
for integer $k$, with $\eta$ given as
\begin{gather}
    \eta \approx \frac{n - 2}{2n}\frac{d}{\mphi}.
\end{gather}
Thus, in the complex $\rho$-plane, the saddles lie close to the imaginary axis. At the saddles, the exponent in \eqref{SDA 3} evaluates to
\begin{gather}
    -n\mphi\l(\ell_{h} + (2k + 1)i\tau_{s}\r) + O(\mphi^{0}),
\end{gather}
where we identify the factor of $\tau_{s} = \frac{\pi}{d}$. Since $\mphi$ is given a negative imaginary part, the leading contribution comes from the \emph{smallest} value of $k$ which can be picked up by a deformation of the integration contour. Around each saddle, the steepest descent of the exponent is in the imaginary $\rho$ direction. From the condition \eqref{SDA 4}, we deduce that $\mcc$ must asymptote close to the negative imaginary axis to define a convergent integral. In this case, starting from the origin we can deform the integration contour along the steepest descent contour in a way that all saddles with $k \geq 0$ are picked up. So, the result derived in \eqref{n point function 4} is indeed correct. \\

\begin{figure}[!ht]
    \centering
    \includegraphics[width=8.5cm]{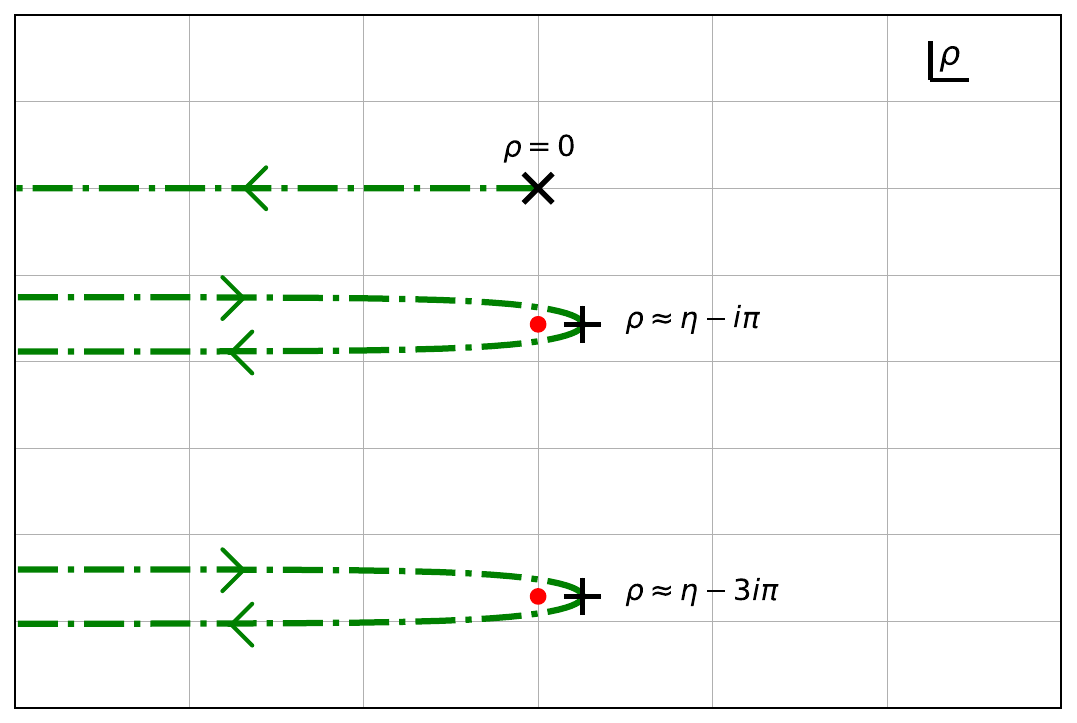}
    \caption{The steepest descent contours for the exponent in \eqref{SDA 3} with $n = 4$.}
    \label{figure: SDA 1}
\end{figure}

The deformed contour passing through the saddles and homologous to $\mcc$ is shown in Figure \ref{figure: SDA 1}. The contours qualitatively appear the same for all $n \geq 3$ and sufficiently large $\mphi$. As discussed for the one-point function in \cite{Grinberg:2020fdj}, there is an end-point contribution to integrals of the kind \eqref{SDA 3} coming from the origin. This comes from integrating over the negative real line in Figure \ref{figure: SDA 1}. However, it is hard to estimate the contribution since WKB expressions are not valid near the origin. The expressions also break down near $\rho \approx 2l\pi$ for integer $l$. This means that for picking up saddles \eqref{SDA 6} with $k \geq 1$, we do not have a rigorous contour deformation argument. Nevertheless, the $k = 0$ saddle is the leading one and our results for the asymptotic form of the $n$-point function still hold. See appendix B of \cite{Grinberg:2020fdj} for additional remarks.

% ---------------------------------------------------------------------

\section{Heavy-light correlator}
\label{Section: heavy light correlator}
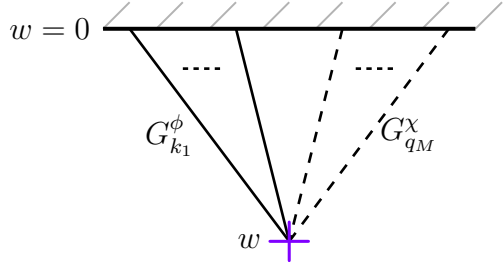
\begin{figure}[!t]
    \centering
    % Witten diagram for n+1 point correlator
    \begin{tikzpicture}[x=1pt,y=1pt,yscale=-1,xscale=1]
    % \path (0,451); %set diagram left start at 0, and has height of 451
    
    % Base boundary line
    \draw [line width = 1.5] (100,110) -- (240,110);
    \draw (80,110) node {$w = 0$};
    
    % Low opacity boundary lines
    \draw [opacity = 0.3, line width = 0.75] (100,110) -- (110,100);
    \draw [opacity = 0.3, line width = 0.75] (120,110) -- (130,100);
    \draw [opacity = 0.3, line width = 0.75] (140,110) -- (150,100);
    \draw [opacity = 0.3, line width = 0.75] (160,110) -- (170,100);
    \draw [opacity = 0.3, line width = 0.75] (180,110) -- (190,100);
    \draw [opacity = 0.3, line width = 0.75] (200,110) -- (210,100);
    \draw [opacity = 0.3, line width = 0.75] (220,110) -- (230,100);
    \draw [opacity = 0.3, line width = 0.75] (240,110) -- (250,100);
    
    % Heavy scalar boundary-bulk propagator
    \draw [line width = 1] (110,110) -- (170,190);
    \draw (125,150) node {$G_{k_{1}}^{\phi}$};
    \draw [dash pattern = on 2pt off 2pt, line width = 1] (130,125) -- (145,125);
    \draw [line width = 1] (150,110) -- (170,190);

    % Light scalar boundary-bulk propagator
    \draw [dash pattern = on 4pt off 4pt, line width = 1] (190,110) -- (170,190);
    \draw [dash pattern = on 2pt off 2pt, line width = 1] (195,125) -- (210,125);
    \draw [dash pattern = on 4pt off 4pt, line width = 1] (230,110) -- (170,190);
    \draw (215,150) node {$G_{q_{M}}^{\chi}$};
    
    % Heavy-light scalars interaction vertex
    \draw (170,190) node [font = \LARGE, color={rgb, 255:red, 128; green, 0; blue, 255}] {$\bm{+}$};
    \draw (155,190) node {$w$};
     
    \end{tikzpicture}

    \caption{Witten diagram for the heavy-light correlator of primaries $\ophi$ and $\ochi$ being sourced holographically by an exchange coupling between the dual scalars. The interaction vertex has been marked and the bulk point is integrated over the exterior region. The $\phi$ propagators are denoted by solid lines while the $\chi$ propagators are denoted by dashed lines. Momenta carried by the propagators are indicated and are small in the sense that they \emph{do not} scale by the mass of the heavy scalar.}
    \label{Witten diagram 2}
\end{figure}

Now, we consider the effect of an exchange interaction between a heavy and a light scalar in the bulk. A coupling between these scalars holographically sources a higher point correlator of the dual primaries. The Witten diagram for such a contribution to the thermal correlator is illustrated in Figure \ref{Witten diagram 2}. The corresponding bulk integral \eqref{Heavy light correlator 1} is performed over the exterior black brane geometry. Plugging in the measure, the integral is easily written in the $w$ variable as follows
\begin{gather}
    \label{Heavy light correlator 2}
    \hspace{-0.2cm} \ev{\ophi(k_{1}) \cdots \ophi(k_{N})\ochi(q_{1}) \cdots \ochi(q_{M})}_{\beta} \sim \kappa_{N,M}\int_{0}^{1}{\frac{dw}{w^{2}}}G_{k_{1}}^{\phi}(w) \ldots G_{k_{N}}^{\phi}(w)\mca_{M}(w),
\end{gather}
where we have clubbed the $\chi$ propagators into
\begin{gather}
    \label{Product of chi propagators 1}
    \mca_{M}(w) \equiv G_{q_{1}}^{\chi}(w) \ldots G_{q_{M}}^{\chi}(w).
\end{gather}

Substituting the WKB form for the $\phi$ propagators, the integral in \eqref{Heavy light correlator 2} reads
\begin{align}
    \begin{split}
        \label{Heavy light correlator 3}
        \ev{\ophi(k_{1}) \cdots \ophi(k_{N})\ochi(q_{1}) \cdots \ochi(q_{M})}_{\beta} & \\
        & \hspace{-5cm} \sim \kappa_{N,M}\int_{0}^{1}{dw}\mca_{M}(w)\exp[-N\mphi\ell_{w} + \frac{N}{4}\log(\frac{w^{2}}{1 - w}) - \log w^{2}].
    \end{split}
\end{align}
The saddle equation for only the exponent in the above integrand will give the leading order behaviour of the integral in the large mass limit. Notice the fact that the exponent in \eqref{Heavy light correlator 3} is same as the one in \eqref{n point function 3} with $n$ replaced by $N$. Of course, this is not a coincidence but simply because the exponent came entirely from the $\phi$ propagator factors. Nonetheless, it should be noted that this occurrence is because we chose not to scale the momenta while deriving the WKB expressions. If instead the momenta were scaled, then the $\chi$ propagators will depend on $\mphi$ and hence contribute to the exponent. \\

The location of the saddle is now simply obtained by replacing $n$ with $N$ in the previously obtained solution \eqref{n point saddle 3}
\begin{gather}
    \label{Heavy light saddle 1}
    w^{*} = -\l(\frac{4N}{N - 8}\frac{\mphi}{d}\r)^{2}.
\end{gather}
Thus, the leading behaviour of the heavy-light thermal correlator is given as
\begin{gather}
    \label{Heavy light correlator 4}
    \ev{\ophi(k_{1}) \cdots \ophi(k_{N})\ochi(q_{1}) \cdots \ochi(q_{M})}_{\beta} \sim \kappa_{N,M}\mca_{M}(w^{*})\exp[-N\mphi(\ell_{h} + i\tau_{s})].
\end{gather}
For a more rigorous justification of this result, we can perform a steepest descent analysis of the integral in \eqref{Heavy light correlator 3}. The computation is identical to the one in section \ref{Section: SDA 1}, so we avoid repeating it. As $\mphi$ is taken to be large, $w^{*}$ (given by \eqref{Heavy light saddle 1}) is also large and hence the factor $\mca_{M}(w^{*})$ can be evaluated by using the asymptotics of the boundary $\chi$ propagator. We suspect that the contribution will not be large, but in case it does modify the exponent, then the shift in the saddle from the $\chi$ propagators should be taken into account.

% ---------------------------------------------------------------------

\section{Higher point correlators and bulk-bulk propagators}
\label{Section: bulk propagator contributions}
The purpose of this section is to discuss higher point thermal correlators which are sourced holographically by Witten diagrams involving bulk-bulk propagators of the heavy scalar. We still look for a signature of time to singularity in higher point thermal correlators while using WKB and saddle point techniques. \\

In \cite{Krishna:2021fus}, the authors considered Witten diagrams involving a bulk-bulk propagator. They approximated the integral by keeping only one of the terms in \eqref{Bulk-bulk propagator 1} for the propagator, thus getting the factor of time to singularity. As we show below, such a restriction is not necessary --- the integrals can be approximated while using the full WKB expression \eqref{Bulk-bulk propagator 2}. \\

The novelty of the calculations and discussion presented can be summarized here itself. Integrals involving bulk-bulk propagators are notorious to evaluate since they typically partition the region of integration. Despite this, we argue for approximating the integral in each region using saddle point method and then summing the contribution. In the correlators we study, the leading behaviour of the integral in each region produces the same exponential factor containing the time to singularity, and hence summing these contributions gives the same leading behaviour for the full correlator. \\

To make things more concrete, in addition to the scalar interactions in \eqref{Action 4}, we add to the action a Weyl squared coupling of the heavy scalar
\begin{gather}
    \label{Action 5}
    S_{\rm W}[\phi] = \alpha\int{d^{d+1}x}\sqrt{g}\,\phi W^{2}.
\end{gather}
In the black brane background, the Weyl tensor squared is given by
\begin{gather}
    \label{Weyl squared 1}
        W^{2} = d(d - 1)^{2}(d - 2)w^{2}.
\end{gather}
This means that for $d \geq 3$, we can form tree level Witten diagrams using a vertex each of \eqref{Action 4} and \eqref{Action 5} (see for example Figures \ref{Witten diagram 3} and \ref{Witten diagram 4}). These contribute, respectively, to corresponding higher point thermal correlators at order $\alpha\lambda_{N}$ and $\alpha\kappa_{N,M}$. \\

Since we work in the probe limit, the Weyl tensor squared in \eqref{Action 5} acts as a bulk source for the $\phi$ field, providing a $w^{2}$ weight for bulk integrals.\footnote{Actually, the $w^{2}$ weight from Weyl tensor squared just cancels against the $\frac{1}{w^{2}}$ coming from the measure.} Furthermore, the momentum conserving delta function at the interaction vertex forces the attached propagator to be at vanishing momentum.

\subsection{\texorpdfstring{$\ev{\ophi(k_{1}) \cdots \ophi(k_{n-1})}_{\beta}$}{$n-1$-point function}}
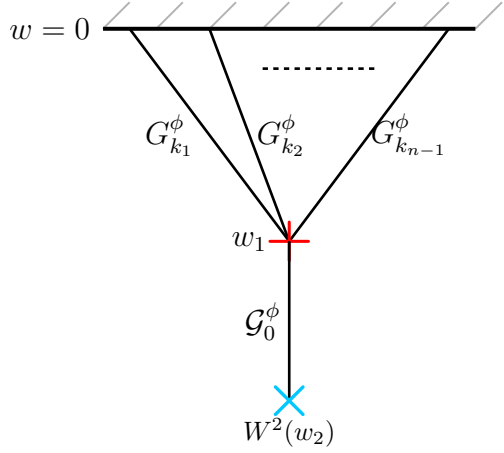
\begin{figure}[!t]
    \centering
    % Witten diagram for higher point heavy-light correlators
    \begin{tikzpicture}[x=1pt,y=1pt,yscale=-1,xscale=1]
    % \path (0,451); %set diagram left start at 0, and has height of 451
    
    % Base boundary line
    \draw [line width = 1.5] (100,110) -- (240,110);
    \draw (80,110) node {$w = 0$};
    
    % Low opacity boundary lines
    \draw [opacity = 0.3, line width = 0.75] (100,110) -- (110,100);
    \draw [opacity = 0.3, line width = 0.75] (120,110) -- (130,100);
    \draw [opacity = 0.3, line width = 0.75] (140,110) -- (150,100);
    \draw [opacity = 0.3, line width = 0.75] (160,110) -- (170,100);
    \draw [opacity = 0.3, line width = 0.75] (180,110) -- (190,100);
    \draw [opacity = 0.3, line width = 0.75] (200,110) -- (210,100);
    \draw [opacity = 0.3, line width = 0.75] (220,110) -- (230,100);
    \draw [opacity = 0.3, line width = 0.75] (240,110) -- (250,100);
    
    % Heavy scalar boundary-bulk propagator
    \draw [line width = 1] (110,110) -- (170,190);
    \draw (125,150) node {$G_{k_{1}}^{\phi}$};
    \draw [line width = 1] (140,110) -- (170,190);
    \draw (167,150) node {$G_{k_{2}}^{\phi}$};
    \draw [dash pattern = on 2pt off 2pt,line width = 1] (160,125) -- (203,125);
    \draw [line width = 1] (230,110) -- (170,190);
    \draw (215,150) node {$G_{k_{n-1}}^{\phi}$};

    % Self interaction vertex
    \draw (170,190) node [font = \LARGE, color={rgb, 255:red, 255; green, 0; blue, 0}] {$\bm{+}$};
    \draw (155,190) node {$w_{1}$};
    
    % Heavy scalar bulk-bulk propagator
    \draw [line width = 1] (170,190) -- (170,249);
    \draw (160,220) node {$\mcg_{0}^{\phi}$};
    
    % Heavy scalar-Weyl squared interaction vertex
    \draw (170,250) node [font = \LARGE, color={rgb, 255:red, 0; green, 200; blue, 255}] {$\bm{\times}$};
    \draw (170,262) node [font = \footnotesize] {$W^{2}(w_{2})$};
    
    \end{tikzpicture}
    
    \caption{Witten diagram for the $n-1$-point function of $\ophi$ being sourced holographically by the self-interaction and the Weyl squared coupling of the dual scalar. The two bulk interaction vertices have been marked separately and both bulk points are integrated over the entire exterior region. The $\phi$ propagators are denoted by solid lines. The momenta carried by the propagators are indicated and are small in the sense that they \emph{do not} scale by the mass of the heavy scalar.}
    \label{Witten diagram 3}
\end{figure}

If we take an $n$-point self-interaction vertex of the heavy scalar along with a vertex of the Weyl squared coupling, then we can source the thermal $n-1$-point function of the dual primary. The Witten diagram is illustrated in Figure \ref{Witten diagram 3}. \\

The bulk integral for the thermal correlator reads as
\begin{gather}
    \label{Bulk self sourced correlator 1}
    \ev{\ophi(k_{1}) \cdots \ophi(k_{n-1})}_{\beta} \sim \alpha\lambda_{n}\int{\frac{dw_{1}dw_{2}}{w_{1}^{2}}}G_{k_{1}}^{\phi}(w_{1}) \ldots G_{k_{n - 1}}^{\phi}(w_{1})\mcg_{0}^{\phi}(w_{1};w_{2}).
\end{gather}
Now, to get the leading exponential behaviour in large $\mphi$ we simply need to evaluate the value of the integrand at the saddle location. Hence, we rewrite the integral as
\begin{gather}
    \label{Bulk self sourced correlator 2}
    \ev{\ophi(k_{1}) \cdots \ophi(k_{n-1})}_{\beta} \sim \alpha\lambda_{n}\int{dw_{1}dw_{2}}\,\mci,
\end{gather}
where the integrand $\mci$ is composed from the propagators and the measure in \eqref{Bulk self sourced correlator 1}. Using the WKB forms \eqref{Boundary-bulk propagator 1} and \eqref{Bulk-bulk propagator 2} for the propagators, the integrand can be expressed as
\begin{gather}
    \label{Bulk self sourced correlator 3}
    \mci = \exp[\mcf(w_{1},w_{2})]
\end{gather}
with the function $\mcf(w_{1},w_{2})$ defined by
\begin{gather}
    \label{Bulk self sourced saddle 1}
    \mcf = -(n - 1)\mphi\ell_{w_{1}} - \mphi\abs{\ell_{w_{1}} - \ell_{w_{2}}} + \frac{1}{4}\log(\frac{w_{1}^{2n - 8}w_{2}^{2}}{(1 - w_{1})^{n}(1 - w_{2})}).
\end{gather}

As we mentioned at the start of the section, the factor of $\abs{\ell_{w_{1}} - \ell_{w_{2}}}$ coming from the bulk-bulk propagator partitions the region of integration into $w_{1} > w_{2}$ and $w_{1} < w_{2}$. In each of these regions, respectively, the replacement $\abs{\ell_{w_{1}} - \ell_{w_{2}}} = \ell_{w_{1}} - \ell_{w_{2}}$ and $\abs{\ell_{w_{1}} - \ell_{w_{2}}} = \ell_{w_{2}} - \ell_{w_{1}}$ has to be made. Thus, the full integrand can be decomposed into two integrands
\begin{gather}
    \label{Bulk self sourced correlator 4}
    \mci = \mci_{a} + \mci_{b}, \\
    \label{Bulk self sourced correlator 5}
    \mci_{a} = \exp[\mcf_{a}(w_{1},w_{2})], \quad \mci_{b} = \exp[\mcf_{b}(w_{1},w_{2})],
\end{gather}
where the functions $\mcf_{a}$ and $\mcf_{b}$ are defined by
\begin{align}
    \label{Bulk self sourced saddle 2}
    \mcf_{a}(w_{1},w_{2}) & = -n\mphi\ell_{w_{1}} + \mphi\ell_{w_{2}} + \frac{1}{4}\log(\frac{w_{1}^{2n - 8}w_{2}^{2}}{(1 - w_{1})^{n}(1 - w_{2})}), \\
    \label{Bulk self sourced saddle 3}
    \mcf_{b}(w_{1},w_{2}) & = -(n - 2)\mphi\ell_{w_{1}} - \mphi\ell_{w_{2}} + \frac{1}{4}\log(\frac{w_{1}^{2n - 8}w_{2}^{2}}{(1 - w_{1})^{n}(1 - w_{2})}).
\end{align}

Note that since each of the integrands $\mci_{a}$ and $\mci_{b}$ has a large parameter ($\mphi$) in the exponent, we can evaluate the corresponding integrals using saddle point.\footnote{The case $n = 2$ is a little different since there is no large parameter accompanying a function of $w_{1}$ in \eqref{Bulk self sourced saddle 3}. Nonetheless, the saddle in $w_{2}$ reproduces the correct expected factor of time to singularity.} Denoting the location of saddle points of $\mcf_{a}(w_{1},w_{2})$ and $\mcf_{b}(w_{1},w_{2})$, respectively, by ($w_{1,a}^{*},w_{2,a}^{*}$) and ($w_{1,b}^{*},w_{2,b}^{*}$), we proceed to solve the equations
\begin{gather}
    \label{Bulk self sourced saddle 4}
    \eval{\p_{w_{1}}\mcf_{a}(w_{1},w_{2})}_{w_{1,a}^{*}, w_{2,a}^{*}} = 0 = \eval{\p_{w_{2}}\mcf_{a}(w_{1},w_{2})}_{w_{1,a}^{*}, w_{2,a}^{*}}, \\
    \label{Bulk self sourced saddle 5}
    \eval{\p_{w_{1}}\mcf_{b}(w_{1},w_{2})}_{w_{1,b}^{*}, w_{2,b}^{*}} = 0 = \eval{\p_{w_{2}}\mcf_{b}(w_{1},w_{2})}_{w_{1,b}^{*}, w_{2,b}^{*}}.
\end{gather}
It is clear from the expressions for $\mcf_{a}$ and $\mcf_{b}$ that the saddle equations for $w_{1}$ and $w_{2}$ decouples for each of them. Substituting for $\mcf_{a}$ and $\mcf_{b}$ and using \eqref{Regularized length 1} for the regularized lengths, the saddle equations reduce to
\begin{align}
    \begin{split}
    \label{Bulk self sourced saddle 6}
    -\frac{n\mphi}{d}\frac{1}{w_{1,a}^{*}\sqrt{1 - w_{1,a}^{*}}} + \frac{2n - 8 - (n - 8)w_{1,a}^{*}}{4w_{1,a}^{*}(1 - w_{1,a}^{*})} & = 0, \\
    \frac{\mphi}{d}\frac{1}{w_{2,a}^{*}\sqrt{1 - w_{2,a}^{*}}} + \frac{2 - w_{2,a}^{*}}{4w_{2,a}^{*}(1 - w_{2,a}^{*})} & = 0,
    \end{split}
    \\
    \begin{split}
    \label{Bulk self sourced saddle 7}
    -\frac{(n - 2)\mphi}{d}\frac{1}{w_{1,b}^{*}\sqrt{1 - w_{1,b}^{*}}} + \frac{2n - 8 - (n - 8)w_{1,b}^{*}}{4w_{1,b}^{*}(1 - w_{1,b}^{*})} & = 0, \\
    -\frac{\mphi}{d}\frac{1}{w_{2,b}^{*}\sqrt{1 - w_{2,b}^{*}}} + \frac{2 - w_{2,b}^{*}}{4w_{2,b}^{*}(1 - w_{2,b}^{*})} & = 0.
    \end{split}
\end{align}

Notice the fact that the equation for $w_{1,a}^{*}$ is exactly the same as the equation for $w^{*}$ in \eqref{n point saddle 2}. This is simply because dependence of $\mcf_{a}$ on $w_{1}$ is same as dependence of the exponent in \eqref{n point function 3} on $w$. The dependence of $\mcf_{b}$ on $w_{1}$ is a little different, thus modifying the saddle equation (and solution) for $w_{1,b}^{*}$ a bit. Also notice that the saddle equations for $w_{2,a}^{*}$ and $w_{2,b}^{*}$ only differ by the sign of the $\mphi$ term. This means that the solution is the same for both in the cut plane but differs in the covering space. \\

The solutions at large $\mphi$ of \eqref{Bulk self sourced saddle 6} and \eqref{Bulk self sourced saddle 7} are given as
\begin{gather}
    \label{Bulk self sourced saddle 8}
    w_{1,a}^{*} \approx -\l(\frac{4n}{n - 8}\frac{\mphi}{d}\r)^{2}, \quad w_{2,a}^{*} \approx -\frac{16\mphi^{2}}{d^{2}}, \\
    \label{Bulk self sourced saddle 9}
    w_{1,b}^{*} \approx -\l(\frac{4(n - 2)}{n - 8}\frac{\mphi}{d}\r)^{2}, \quad w_{2,b}^{*} \approx -\frac{16\mphi^{2}}{d^{2}}.
\end{gather}
Now, we simply need to evaluate the integrands $\mci_{a}$ and $\mci_{b}$ at these saddle values. Using the same prescription as before, we evaluate the regularized lengths at the saddles as
\begin{gather}
    \label{Regularized length 3}
    \ell_{w_{1,a}^{*}} = \ell_{w_{2,a}^{*}} = \ell_{w_{1,b}^{*}} = \ell_{w_{2,b}^{*}} = \ell_{h} + i\frac{\pi}{d} + O\l(\frac{1}{\mphi}\r).
\end{gather}
so that each of the integrands start of as 
\begin{gather}
    \label{Bulk self sourced correlator 6}
    \mci_{a} \sim \exp[-(n - 1)\mphi(\ell_{h} + i\tau_{s})], \\
    \label{Bulk self sourced correlator 7}
    \mci_{b} \sim \exp[-(n - 1)\mphi(\ell_{h} + i\tau_{s})].
\end{gather}
Thus, we conclude that the leading behaviour of the contribution coming from Witten diagram in Figure \ref{Witten diagram 3} to the thermal $n-1$-point function of $\ophi$ is given as
\begin{gather}
    \label{Bulk self sourced correlator 8}
    \ev{\ophi(k_{1}) \cdots \ophi(k_{n - 1})}_{\beta} \sim \alpha\lambda_{n}\exp[-(n - 1)\mphi(\ell_{h} + i\tau_{s})].
\end{gather}

\subsection{\texorpdfstring{$\ev{\ophi(k_{1}) \cdots \ophi(k_{N-1})\ochi(q_{1}) \cdots \ochi(q_{M})}_{\beta}$}{Heavy-light correlator}}
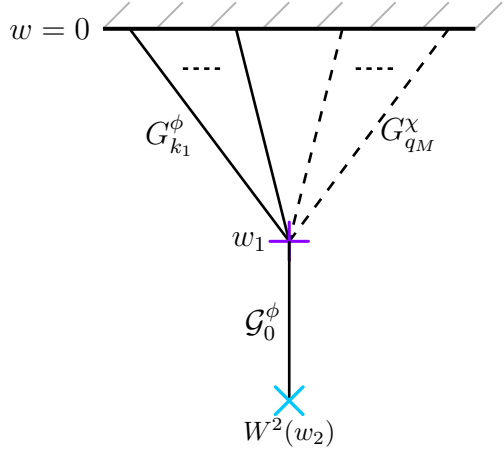
\begin{figure}[!t]
    \centering
    % Witten diagram for higher point heavy-light correlators
    \begin{tikzpicture}[x=1pt,y=1pt,yscale=-1,xscale=1]
    % \path (0,451); %set diagram left start at 0, and has height of 451
    
    % Base boundary line
    \draw [line width = 1.5] (100,110) -- (240,110);
    \draw (80,110) node {$w = 0$};
    
    % Low opacity boundary lines
    \draw [opacity = 0.3, line width = 0.75] (100,110) -- (110,100);
    \draw [opacity = 0.3, line width = 0.75] (120,110) -- (130,100);
    \draw [opacity = 0.3, line width = 0.75] (140,110) -- (150,100);
    \draw [opacity = 0.3, line width = 0.75] (160,110) -- (170,100);
    \draw [opacity = 0.3, line width = 0.75] (180,110) -- (190,100);
    \draw [opacity = 0.3, line width = 0.75] (200,110) -- (210,100);
    \draw [opacity = 0.3, line width = 0.75] (220,110) -- (230,100);
    \draw [opacity = 0.3, line width = 0.75] (240,110) -- (250,100);
    
    % Heavy scalar boundary-bulk propagator
    \draw [line width = 1] (110,110) -- (170,190);
    \draw (125,150) node {$G_{k_{1}}^{\phi}$};
    \draw [dash pattern = on 2pt off 2pt, line width = 1] (130,125) -- (145,125);
    \draw [line width = 1] (150,110) -- (170,190);

    % Light scalar boundary-bulk propagator
    \draw [dash pattern = on 4pt off 4pt, line width = 1] (190,110) -- (170,190);
    \draw [dash pattern = on 2pt off 2pt, line width = 1] (195,125) -- (210,125);
    \draw [dash pattern = on 4pt off 4pt, line width = 1] (230,110) -- (170,190);
    \draw (215,150) node {$G_{q_{M}}^{\chi}$};
    
    % Heavy-light scalars interaction vertex
    \draw (170,190) node [font = \LARGE, color={rgb, 255:red, 138; green, 0; blue, 255}] {$\bm{+}$};
    \draw (155,190) node {$w_{1}$};
    
    % Heavy scalar bulk-bulk propagator
    \draw [line width = 1] (170,190) -- (170,249);
    \draw (160,220) node {$\mcg_{0}^{\phi}$};
    
    % Heavy scalar-Weyl squared interaction vertex
    \draw (170,250) node [font = \LARGE, color={rgb, 255:red, 0; green, 200; blue, 255}] {$\bm{\times}$};
    \draw (170,262) node [font = \footnotesize] {$W^{2}(w_{2})$};
    
    \end{tikzpicture}
    
    \caption{Witten diagram for heavy-light correlator being sourced holographically by bulk couplings between the dual fields and the Weyl squared coupling of the heavy scalar. The two bulk interaction vertices have been marked separately and both the bulk points are integrated over the entire exterior region. Solid lines denote propagators for the heavy scalar, $\phi$, and dashed ones are for the light scalar, $\chi$. The momenta carried by the propagators are indicated and are small in the sense that they \emph{do not} scale by the mass of the heavy scalar.}
    \label{Witten diagram 4}
\end{figure}

Similar to the previous computation, we now consider a contribution to a heavy-light thermal correlator coming from the Witten diagram formed from a heavy-light vertex and a Weyl squared interaction vertex. The diagram is shown in Figure \ref{Witten diagram 4}. \\

The bulk integral for the contribution reads as
\begin{align}
    \begin{split}
        \label{Bulk heavy light correlator 1}
        \ev{\ophi(k_{1}) \cdots \ophi(k_{N-1})\ochi(q_{1}) \cdots \ochi(q_{M})}_{\beta} & \\
        & \hspace{-5cm} \sim \alpha\kappa_{N,M}\int{\frac{dw_{1}dw_{2}}{w_{1}^{2}}}G_{k_{1}}^{\phi}(w_{1}) \ldots G_{k_{N-1}}^{\phi}(w_{1})\mcg_{0}^{\phi}(w_{1};w_{2})\mca_{M}(w_{1}),
    \end{split}
\end{align}
where $\mca_{M}(w)$ is the same product of propagators defined in \eqref{Product of chi propagators 1}. The product of $\phi$ propagators is the same as that appearing in the integrand of \eqref{Bulk self sourced saddle 1} with the replacement of $n$ with $N$. Since the factor of $\mca_{M}(w_{1})$ does not contribute any terms with the large parameter $\mphi$, the leading behaviour from the saddle point analysis can be obtained by considering the remaining integrand only. Hence, the results of the previous subsection can be borrowed. \\

Thus, we deduce that the contribution to the thermal correlator \eqref{Bulk heavy light correlator 1} is approximated to leading order as
\begin{align}
    \begin{split}
        \label{Bulk heavy light correlator 2}
        \ev{\ophi(k_{1}) \cdots \ophi(k_{N-1})\ochi(q_{1}) \cdots \ochi(q_{M})}_{\beta} & \\
        & \hspace{-6cm} \sim \alpha\kappa_{N,M}\exp[-(N - 1)\mphi(\ell_{h} + i\tau_{s})]\l(c_{a}\mca_{M}(w_{1,a}^{*}) + c_{b}\mca_{M}(w_{1,b}^{*})\r).
    \end{split}
\end{align}
Here, $w_{1,a}^{*}$ and $w_{1,b}^{*}$ are obtained, respectively, from \eqref{Bulk self sourced saddle 8} and \eqref{Bulk self sourced saddle 9} by replacing $n$ with $N$. $c_{a}$ and $c_{b}$ are the coefficients coming from performing the integral after the partitioning due to the bulk-bulk propagator. These will, at most, contain powers of $\mphi$.

% ---------------------------------------------------------------------
{\color{black}
\section{Other symmetric solutions}
\label{Section: other symmetric solutions}
So far, we have established that factors of proper time to singularity in the black brane background are present in very generic contributions to higher point thermal CFT correlators of a heavy primary operator. The dual CFT in this case is defined on $S_{\beta}^{1} \times \bbr^{d - 1}$. By changing the bulk black hole background it is possible to consider the CFT on a different geometry as well. Conversely, to probe the interior structure of different black hole spacetimes we need to study CFTs defined on different geometries. \\

In this section, we consider two $d + 1$-dimensional black hole spacetimes: the hyperbolic black hole and the charged black brane. The dual CFT in the first case is defined on the hyperbolic space $S_{\beta}^{1} \times H^{d - 1}$ while the dual to the second case is a CFT on $S_{\beta}^{1} \times \bbr^{d - 1}$ with a chemical potential. The $n$-point self-interaction term in \eqref{Action 4} sources the thermal $n$-point function of the dual operator in the corresponding CFT via the Witten diagram in Figure \ref{Witten diagram 1}. In the large mass limit, we use WKB methods to approximate the relevant boundary-bulk propagators in the two black hole backgrounds and evaluate the respective bulk integral using saddle point method. The result for the $n$-point functions in the two backgrounds is shown in \eqref{Hyperbolic n point function 2} and \eqref{Charged BB n point function 2}. Thus, we see that in both the cases the geometrical lengths in the black hole interiors are probed by the higher point thermal correlators of primaries of large conformal dimension. \\

In a way, this result is expected since the WKB and saddle techniques used in the case of the black brane easily generalize to other backgrounds. In particular, the radial geodesic distance to a bulk point from the boundary is the first term in the WKB limit of the boundary-bulk propagator. Further, the saddle for the bulk integrals lie `close' to singularity (in the full complex space) so that saddle point answer naturally contains geometrical lengths associated to the singularity. This was already demonstrated in \cite{Grinberg:2020fdj} and \cite{David:2022nfn} by the computation of thermal one-point function in different spherical backgrounds. We have generalized the result for thermal $n$-point functions. The extension to heavy-light correlators and diagrams with bulk-bulk propagators is straightforward.

\subsection{Hyperbolic black holes}
We repeat here the calculation of section \ref{Section: n point function} for the case of hyperbolic black holes \cite{Birmingham:1998nr}. David and Kumar \cite{David:2022nfn} studied the Gauss-Bonnet coupling of a heavy scalar in the hyperbolic black hole background. Similar to the Weyl squared term, this coupling sources the thermal one point function in the CFT defined on the hyperbolic boundary. By performing the bulk integral exactly, it was shown in \cite{David:2022nfn} that the thermal one point function in the limit of large conformal dimension does contain a factor of proper time to singularity in the exponent. Similar to the calculation in section \ref{Section: n point function}, we now show that even in the hyperbolic black hole background a self interaction term for a heavy scalar gives factors of time to singularity in the exponent of the $n$-point thermal correlator in the dual CFT. \\

The Euclidean hyperbolic black hole metric in $d + 1$-dimensions is given as
\begin{gather}
    \label{Hyperbolic bh metric 1}
    ds^{2} = \frac{R^{2}}{z^{2}}\l(f(z)dt^{2} + \frac{1}{f(z)}dr^{2} + R^{2}(du^{2} + \sinh^{2}u \,d\O_{d - 2}^{2}) \r) \quad ; \quad f(z) = 1 - \frac{z^{2}}{R^{2}}.
\end{gather}
We keep the AdS length scale, $R$, explicit since the horizon is at $z = R$. The proper radial time to singularity in the above background is given as
\begin{gather}
    \label{Hyperbolic time to singularity 1}
    \tau_{s} = R\int_{R}^{\infty}{dz}\frac{1}{z\sqrt{-f(z)}} = \frac{\pi}{2}R.
\end{gather}
Since the metric is highly isometric, solutions of the Klein-Gordon equation in this background can be decomposed as
\begin{gather}
    \label{Hyperbolic modes 1}
    \phi(t,z,u,\O) = \sum_{lm}\int{d\o}\int{d\lambda}\,\phi_{\sigma}(z)e^{-i\o t}q_{\lambda l}(u)Y_{lm}(\O),
\end{gather}
with $\sigma \equiv (\o,\lambda,l,m)$. Here, $Y_{lm}(\O)$ are the usual spherical harmonics on $S^{d - 2}$ and $\lambda \in [0,\infty)$ is a continuous parameter labelling the function $q_{\lambda,l}(u)$ defined by (see \cite{Camporesi:1994ga,David:2022nfn})
\begin{gather}
    \label{Hyperbolic modes 2}
    q_{\lambda l}(u) = (i\sinh u)^{l}\hf\l(\xi + i\lambda,\xi - i\lambda;\frac{1}{2} + \xi;-\sinh^{2}\frac{u}{2} \r) \quad ; \quad \xi = \frac{d - 2}{2} + l.
\end{gather}
Feeding \eqref{Hyperbolic modes 1} into the Klein-Gordon equation gives the following radial equation for the modes $\phi_{\sigma}(z)$
\begin{gather}
    \label{Hyperbolic modes 3}
    \frac{z^{d + 1}}{R^{d + 1}}\p_{z}\l(\frac{R^{d - 1}}{z^{d - 1}}f(z)\p_{z}\phi_{\sigma}(z)\r) - \l(\frac{\o^{2}}{f(z)}\frac{z^{2}}{R^{2}} + \l(\lambda^{2} + \l(\frac{d - 2}{2}\r)^{2}\r)\frac{z^{2}}{R^{4}} + \mphi^{2}\r)\phi_{\sigma}(z) = 0.
\end{gather}

In the large mass limit, we solve equation \eqref{Hyperbolic modes 3} using a WKB ansatz $\phi_{\sigma}(z) \propto \exp(\mphi S(z))$, where $S(z)$ is expanded in powers of $\frac{1}{\mphi}$. The leading order WKB solution does not involve the parameters $\sigma$.\footnote{\color{black} This is because the parameters are held fixed while taking the large mass limit.} Substituting the ansatz into \eqref{Hyperbolic modes 3}, we find solutions for $S(z)$. These are conveniently specified in the variable $w = \frac{z^{2}}{R^{2}}$ as
\begin{gather}
    \label{Hyperbolic WKB 1}
    S(z) \equiv S(w) = \pm \ell_{w} + \frac{1}{4\mphi}\log(\frac{w^{d}}{1 - w}) + O\l(\frac{1}{\mphi^{2}}\r).
\end{gather}
The $\pm$ sign corresponds to the normalizable and non-normalizable solution, respectively. The regularized radial geodesic distance $\ell_{w}$ is given by the integral
\begin{gather}
    \label{Hyperbolic geodesic distance 1}
    \ell_{w} = \frac{R}{2}\lim_{w_{c} \to 0}\l(\int_{w_{c}}^{w}\frac{dw}{w\sqrt{1 - w}} + \log w_{c}\r) = \ell_{h} - R\tanh^{-1}\sqrt{1 - w}
\end{gather}
where the identification $\ell_{h} \equiv R\log2$ has been made. Notice the similarity between the expressions obtained here and in section \ref{Section: preliminaries} for the black brane. We require the boundary-bulk propagator which is simply proportional to the non-normalizable solution and is given as
\begin{gather}
    \label{Hyperbolic boundary bulk propagator 1}
    G_{\sigma}^{\phi}(w) \sim \exp[-\mphi\ell_{w} + \frac{1}{4}\log(\frac{w^{d}}{1 - w})] \times \text{powers of $\mphi$}
\end{gather}

Now, assuming the self-interaction $\frac{\lambda_{n}}{n!}\phi^{n}$ in the hyperbolic black hole background, the dual thermal $n$-point function for the operators $\ophi$ is sourced by the contact Witten diagram shown in Figure \ref{Witten diagram 1}. Using the WKB propagators \eqref{Hyperbolic boundary bulk propagator 1} we write the the bulk integral contributing to the thermal correlator as\footnote{\color{black} The CFT operators have also been decomposed into the modes on the hyperboloid and hence are characterized by the labels $\sigma_{i}$. An overall delta function $\delta^{(d)}(\sum_{i}\sigma_{i})$ coming from integrals in transverse directions has been suppressed.}
\begin{gather}
    \label{Hyperbolic n point function 1}
    \ev{\ophi(\sigma_{1})\cdots\ophi(\sigma_{n})}_{\beta} \sim \lambda_{n}\int_{0}^{1}{dw}\exp[-n\mphi\ell_{w} + \frac{n}{4}\log(\frac{w^{d}}{1 - w}) - \l(\frac{d}{2} + 1\r)\log w].
\end{gather}
The saddle equation for the above exponent reads as
\begin{gather}
    \label{Hyperbolic saddle 1}
    -\frac{n\mphi R}{2w^{*}\sqrt{1 - w^{*}}} + \frac{nd - 2d - 4 - (nd - 2d - n - 4)w^{*}}{4w^{*}(1 - w^{*})} = 0.
\end{gather}
For large $\mphi$, the solution is
\begin{gather}
    \label{Hyperbolic saddle 2}
    w^{*} \approx -\l(\frac{2n\mphi R}{d + 5}\r)^{2}.
\end{gather}
Note that as $\mphi \to \infty$ the saddle lies near the singularity. The regularized length at the saddle is defined by giving $\mphi$ a negative imaginary part, so that putting the saddle value \eqref{Hyperbolic saddle 2} into \eqref{Hyperbolic geodesic distance 1} gives
\begin{gather}
    \label{Hyperbolic geodesic distance 2}
    \ell_{w^{*}} \approx \ell_{h} + i\tau_{s},
\end{gather}
where we have identified the factor of proper time to singularity obtained in \eqref{Hyperbolic time to singularity 1}. Finally, at leading order, the thermal correlator \eqref{Hyperbolic n point function 1} is given by the value of the integrand at the saddle point. We thus get the desired form for the thermal $n$-point function
\begin{gather}
    \label{Hyperbolic n point function 2}
    \ev{\ophi(\sigma_{1})\cdots\ophi(\sigma_{n})}_{\beta} \sim \lambda_{n}\exp[-n\mphi(\ell_{h} + i\tau_{s})] \times \text{powers of $\mphi$}.
\end{gather}

\subsection{Charged black branes}
We now consider a charged black hole background. The interior of charged black holes is qualitatively different from that of uncharged ones --- there is an inner horizon and the singularity is time-like. An infalling observer on a radial geodesic from the outer horizon takes a finite proper time ($\tau_{in}$) to reach the inner horizon but avoids falling into the singularity. In contrast, a space-like radial geodesic starting at the inner horizon meets the singularity at a finite proper distance ($\ell_{sing}$). Following the calculation of section \ref{Section: n point function}, we demonstrate that a similar bulk computation of the $n$-point function in the charged background features both $\tau_{in}$ and $\ell_{sing}$, apart from the horizon distance $\ell_{h}$. \\

We start by taking the charged black brane in \adsd.\footnote{\color{black} We can also consider a spherically symmetric black hole instead of a black brane. The only difference is that the transverse modes are spherical harmonics instead of plane waves. This does not matter in the computation since only the purely radial modes are important in the large mass limit.} The metric takes the same form as \eqref{Black brane metric 1} but with a different emblackening factor $f(z)$
\begin{gather}
    \label{Charged BB metric 1}
    ds^{2} = \frac{1}{z^{2}}\l(f(z)dt^{2} + \frac{1}{f(z)}dz^{2} + d\vec{x}^{2} \r) \quad ; \quad f(z) = 1 - \frac{z^{d}}{z_{0}^{d}} + qz^{2d - 2}.
\end{gather}
The AdS length scale has been set to unity. The Klein-Gordon equation in terms of the Fourier modes reduces to the radial equation given in \eqref{KG equation 2} but with $f(z)$ specified in \eqref{Charged BB metric 1}. In the large mass limit, we substitute the WKB ansatz $\phi_{k} \propto \exp(\mphi S(z))$ with $S(z)$ admitting an expansion in $\frac{1}{\mphi}$. The leading order solutions are readily obtained as
\begin{gather}
    \label{Charge BB WKB 1}
    S(z) = \pm \ell_{z} + \frac{1}{4\mphi}\log(\frac{z^{2d}}{z_{0}^{2d}f(z)}) + O\l(\frac{1}{\mphi^{2}}\r).
\end{gather}
As before, $\ell_{z}$ is the proper distance to a bulk point $z$ from the boundary and $\pm$ corresponds to the normalizable and non-normalizable solutions. Explicitly, we define
\begin{gather}
    \label{Charged BB geodesic distance 1}
    \ell_{z} = \lim_{z_{c} \to 0}\l(\int_{z_{c}}^{z}\frac{dz}{z\sqrt{f(z)}} + \log(\frac{z_{c}}{z_{0}})\r).
\end{gather}
The branch cut for the integral is taken between the outer and inner horizons. The proper length to a bulk point (possibly in the black hole interior) depends on which sheet the point is taken as well as the specific contour. It will be important for us to calculate the proper length to the singularity, $z \to \infty$. There are (at least) four choices of contours corresponding to whether it lies entirely on the principle branch or goes to the second sheet as well as the choice of whether it goes over the branch cut or below it. The proper lengths for these choices are succinctly expressed as
\begin{gather}
    \label{Charged BB geodesic distance 2}
    \ell_{z \to \infty} = \ell_{h} \pm i\tau_{in} \pm\ell_{sing},
\end{gather}
where the two $\pm$ are independently chosen, thus leading to four possibilities.\footnote{\color{black} There are actually more possibilities coming from the fact that the contour can wind around infinity arbitrary number of times. The contribution to the proper length from such windings is purely imaginary. Further, there may be more branch points and cuts for the integral in \eqref{Charged BB geodesic distance 1} which gives more contours.} In the above expression, the horizon length is defined as
\begin{gather}
    \label{Charged BB geodesic distance 3}
    \ell_{h} = \lim_{z_{c} \to 0}\l(\int_{z_{c}}^{z_{out}}\frac{dz}{z\sqrt{f(z)}} + \log(\frac{z_{c}}{z_{0}})\r),
\end{gather}
while the time between the two horizons and the proper length between the inner horizon and singularity is given by the integrals
\begin{gather}
    \label{Charged BB geodesic distance 4}
    \tau_{in} = \int_{z_{out}}^{z_{in}}\frac{dz}{z\sqrt{-f(z)}}, \quad \ell_{sing} = \int_{z_{in}}^{\infty}\frac{dz}{z\sqrt{f(z)}}.
\end{gather}

We can now compute the Witten diagram in Figure \ref{Witten diagram 1} contributing to the thermal $n$-point function, but for the background in \eqref{Charged BB metric 1}. As before, we make use of the WKB expressions for the boundary-bulk propagators in the bulk integral. After simple substitution, the integral for the $n$-point function reads as (suppressing the delta function for the conservation of transverse momentum)
\begin{gather}
    \label{Charged BB n point function 1}
    \ev{\ophi(k_{1}) \cdots \ophi(k_{n})}_{\beta} \sim \lambda_{n}\int_{0}^{z_{out}}{dz}\exp[-n\mphi\ell_{z} + \frac{n}{4}\log(\frac{z^{2d}}{z_{0}^{2d}f(z)}) - \log z^{d + 1}].
\end{gather}
The saddle equation for the exponent is
\begin{gather}
    \label{Charged BB saddle 1}
    -\frac{n\mphi}{z^{*}\sqrt{f(z^{*})}} - \frac{n}{4}\frac{f'(z^{*})}{f(z^{*})} + \frac{nd - 2d - 2}{2z^{*}} = 0.
\end{gather}
At large $\mphi$ and finite $q$, the solution is approximated by
\begin{gather}
    \label{Charged BB saddle 2}
    (z^{*})^{d - 1} \approx \frac{2n\mphi}{(n - 2 - 2d)q^{2}}.
\end{gather}
Again, as $\mphi \to \infty$ the saddle is near the singularity. Hence, at leading order the $n$-point function is just given by evaluating the integrand at the saddle. As already discussed, depending on the branch on which the saddle is taken we get one of the answers in \eqref{Charged BB geodesic distance 2}. As with the other black holes, the bulk integral has to be defined by analytic continuation. We do this by giving $\mphi$ a negative imaginary part. This forces us to pick a contour lying \emph{above} the branch cut while evaluating the proper length to singularity, thus giving a contribution of $i\tau_{in}$. In general, we will pick up contributions from saddles on both branches so that both signs of $\ell_{sing}$ contribute. Since it multiplies $-n\mphi$, the leading answer comes from the saddle on the second branch, i.e. by picking $-\ell_{sing}$. So, we get the leading contribution by putting in the following value for the proper length to singularity
\begin{gather}
    \label{Charge BB geodesic distance 5}
    \ell_{z^{*}} \approx \ell_{h} + i\tau_{in} - \ell_{sing}.
\end{gather}
We thus get the following form for the $n$-point function
\begin{gather}
    \label{Charged BB n point function 2}
    \ev{\ophi(k_{1}) \cdots \ophi(k_{n})}_{\beta} \sim \lambda_{n}\exp[-n\mphi(\ell_{h} + i\tau_{in} - \ell_{sing})] \times \text{powers of $\mphi$}.
\end{gather}
}

% ---------------------------------------------------------------------

\section{Conclusion}
We have argued in this paper that for a thermal CFT admitting a holographic description of probe fields in a black brane background, certain higher point thermal correlators may contain the information of the black hole singularity. We consider the bulk theory with a self-interaction term of a heavy scalar and a coupling between a heavy and light scalar. \\

The interactions source thermal correlators of the dual CFT primaries, which are given by bulk integrals of propagators over the Euclidean geometry. In the limit of large conformal dimension, or equivalently large mass of the heavy scalar, these integrals can be evaluated using WKB forms of the propagators and saddle point method. We have shown that the saddle contribution contains geometrical factors such as the proper radial time to singularity from the horizon. The explicit form of the thermal correlators is given by \eqref{n point function 4} and \eqref{Heavy light correlator 4}. The use of WKB and saddle point approximations is itself justified by its success in matching with the exact calculation of the thermal one point function being sourced by a bulk coupling of the scalar to the Weyl tensor squared \cite{Grinberg:2020fdj}. This has been reviewed in our notation in appendix \ref{Section: one point function}. {\color{black} To provide further support for WKB methods, in appendix \ref{Section: deriving WKB expressions} we use known exact solutions to derive WKB expressions for the zero momentum boundary-bulk propagator in the black brane as well as the finite momentum boundary-bulk propagator in the BTZ black hole.} \\

% Factorization property?

An important conclusion that we emphasize is that to obtain a signature of singularity in thermal CFT correlators, it was not necessary to include couplings of dual scalars to higher derivative curvature terms. Simple self-interaction terms also work to give information of the time to singularity. The reason is that the saddle of the bulk integrals lies at $\infty$ (in the $w$ variable), and then the WKB expressions have the proper length to the bulk vertex in the exponent, thus giving a factor of the time to singularity. Such computations \emph{do not} necessitate factors coming from curvature terms. \\

We have also considered contributions to thermal correlators coming from Witten diagrams containing bulk-bulk propagators. Bulk-bulk propagators partition the region of integration making the analysis a bit involved. Nonetheless, it is argued that each integral can be approximated to leading order using saddle point method. Since each of them contains the same exponent containing the time to singularity, the full contribution of the Witten diagram also has the same form. \\

The temperature of the CFT is inversely proportional to the horizon radius, $T \propto z_{h}$. So, by restoring factors of $z_{h}$ in the propagators and the measure, we can recover temperature dependence of the correlators. For instance, the overall temperature dependence in \eqref{n point function 4} of the $n$-point function of $\ophi$ can be fixed to $(T)^{n(\Delta_{\phi} - d) + d}$, since a factor of $z_{h}^{d - \Delta_{\phi}}$ comes from each propagator insertion (see for example \cite{David:2022nfn} for fixing normalization of the propagators), and $z_{h}^{-d}$ comes from the measure of the bulk integral. Similarly, the temperature dependence of the heavy-light correlator in \eqref{Heavy light correlator 4} can be fixed as $T^{N(\Delta_{\phi} - d) + M(\Delta_{\chi} - d) + d}$. If we evaluate the next order terms in the WKB expansion of the propagator, both momentum dependence and factors of $z_{h}$ will appear. Thus, the terms subleading in $\frac{1}{\mphi}$ have a different temperature dependence than the leading term. \\

A significant advantage of the WKB approach over just using the geodesic approximations, as in \cite{Rodriguez-Gomez:2021pfh,Rodriguez-Gomez:2021mkk,Georgiou:2022ekc,Georgiou:2023xpg}, to get the form of the propagators is that the corrections coming from the momentum and the temperature dependence can be computed systematically even at small values of the momentum. At large momentum (scaled by $\mphi$), the expressions for the propagators are more complicated and it is not clear whether the saddle analysis will give information about the singularity. Nonetheless, the WKB expressions can be corrected and are useful for computing bulk integrals. \\

It is evident from David and Kumar's analysis \cite{David:2022nfn} that the structure of the thermal one point function in \eqref{One point function 1} holds for more general (`spherically' symmetric) black holes. {\color{black} We have demonstrated in section \ref{Section: other symmetric solutions} that geometrical lengths corresponding to interiors of more general black holes appear in higher point thermal correlators as well.} In highly isometric black hole solutions, the wave equation reduces to an ODE in a single (radial) variable, with the coefficients containing parameters corresponding to the isometries (analogue of $k = (\o,\vec{k})$). In such a case, it is generically possible to use the WKB ansatz to find a solution in a $\frac{1}{\mphi}$ expansion. If in the large $\mphi$ limit other parameters are not scaled, then we expect the saddle arguments to go through. The precise location of the saddle in the radial variable may, of course, differ. But as $\mphi$ is taken to $\infty$, the saddle {\color{black} also goes} to $\infty$ and {\color{black} produces} the factor of time to singularity, or for black holes with inner horizon, the factor of time to inner horizon from the outer one and space-like distance to singularity from the inner horizon. \\

Although our computation of the thermal correlators is holographic and relies on the supergravity approximation of the bulk theory, it is general enough that simple bulk interaction terms make the correlators vary exponentially with the dimension of the fields. It is interesting to consider whether this behaviour is preserved when the 't Hooft coupling is not large, i.e. when supergravity is no longer a good approximation. In fact, it may be the case that such a variation of thermal correlators of heavy fields is a general feature in weakly coupled QFTs as well. One can then rely on perturbative or non-perturbative methods at weak coupling for verifying a feature like this directly from the QFT side. This should be a less formidable task than trying to verifying this feature in strongly coupled CFTs as attempted in \cite{David:2023uya}. \\

There are several ways for improving upon the analysis done here. To start with, obtaining the form of the corrections coming from the higher order WKB terms containing factors of the momentum will give us more information about the dynamics in the thermal CFT. In particular, it may be possible to connect the factor of time to singularity appearing as a `phase' in the higher point thermal correlators with diffusive phenomena in the CFT. On the other hand, the method followed here may be helpful in explicitly performing bulk calculations in black hole backgrounds which do not have spherical symmetry, for example rotating black holes.\footnote{I thank Justin David and Srijan Kumar for suggesting this possibility.} It will be nice if the method presented in this paper can be generalized to thermal correlation functions of primaries carrying large momentum (scaling with the conformal dimension) to see whether these contain geometrical features of the bulk geometry, and in particular, of the black hole. \\

Lastly, we conclude with a question regarding the generalization of the entire proposal --- reading off the proper time to singularity from CFT correlators --- to dynamical backgrounds.\footnote{I thank Suvrat Raju for putting forth this very legitimate concern so succinctly.} In a time dependent black hole geometry, the time to singularity is a function of when the observer falls into the black hole. The CFT dual to a dynamical black hole geometry is an out-of-equilibrium state. The general expectation then, from holography, is that correlators in this state (in some appropriate WKB limit) should contain the different factors of time to singularity. These correlators can be computed using real time holographic methods (for example, see \cite{Jana:2020vyx}) and there is ongoing work in this direction for collapse geometries.

\acknowledgments
I express my gratitude towards my mentor, R. Loganayagam, for helping me improve upon the draft and painfully going over a few calculations with me. I would like to thank Justin David and Srijan Kumar for illuminating discussions. I also thank Gautam Mandal, Godwin Martin, Suvrat Raju, Ashoke Sen, Shivam Sharma, Omkar Shetye, Shridhar Vinayak, Avi Wadhwa, and the rest of the ICTS string group for various interactions resulting in an improvement of this work. I acknowledge support of the Department of Atomic Energy, Government of India, under project no. RTI4001.

% ---------------------------------------------------------------------
\appendix

{\color{black}
\section{Deriving WKB approximations from exact expressions}
\label{Section: deriving WKB expressions}
In the bulk of the paper we have used WKB expressions for propagators and saddle point methods for bulk integrals to compute holographic contributions to higher point thermal correlators of large conformal dimension primaries. It will be nice to develop other checks for justifying this approach. An example is the original problem of Grinberg and Maldacena \cite{Grinberg:2020fdj} where they compute the thermal one point function exactly in the presence of a bulk Weyl squared coupling and show agreement with the WKB approach (see appendix \ref{Section: one point function} for a quick review of their calculation). A similar check via exact computation for higher point functions does not seem to be possible due to the numerous insertions of boundary-bulk propagators in the bulk integrals. \\

Instead, we argue that in cases where the exact propagator is available, the corresponding WKB expression can be derived using an appropriate integral representation and method of steepest descent. Then, the computation of bulk integrals with many propagators can be thought of as successive approximations, thus justifying our approach.
}

% ---------------------------------------------------------------------

\subsection{WKB expression of \texorpdfstring{$G_{0}^{\phi}$}{boundary-bulk propagator at vanishing momentum}}
\label{Section: consistency check 1}
{\color{black} Here, we} provide an alternative derivation for the WKB expression \eqref{Boundary-bulk propagator 1} of the boundary-bulk propagator $G_{0}^{\phi}(w)$ at vanishing momentum. The exact solution is given as (suppressing overall factors that do not contribute to the exponent in large $\hphi$ limit) \cite{David:2022nfn}
\begin{gather}
    \label{Boundary-bulk propagator 2}
    G_{0}^{\phi}(w) = \frac{\G(\hphi)^{2}}{\G(2\hphi)}w^{\hphi}\hf(\hphi,\hphi;1;1 - w).
\end{gather}

We make use of the following integral representation for the hypergeometric (equation 9.112 in \cite{tableofintegrals})
\begin{gather}
    \label{Integral representation 1}
    \hf(p,p + n;n + 1;z^{2}) = z^{-n}\frac{\G(p)n!}{\G(p + n)}\int_{0}^{2\pi}{\frac{dt}{2\pi}}\frac{\cos nt}{(1 - 2z\cos t + z^{2})^{p}},
\end{gather}
where $n$ is a whole number and $p \neq 0, -1, -2, \ldots$. The hypergeometric appearing in \eqref{Boundary-bulk propagator 2} is obtained by putting $n = 0$, $p = \hphi$ and $z = \sqrt{1 - w}$. The expression for the propagator then reads
\begin{gather}
    \label{Boundary-bulk propagator 3}
    G_{0}^{\phi}(w) = \frac{\G(\hphi)^{2}}{\G(2\hphi)}w^{\hphi}\int_{0}^{2\pi}{\frac{dt}{2\pi}}\frac{1}{\l(2 - w - 2\sqrt{1 - w}\cos t \r)^{\hphi}}.
\end{gather}
For large $\hphi$, the $t$ integral can be performed using saddle point. It is easy to see that saddles lie at solutions of $\sin t = 0$, i.e. at $t = n\pi$ for integers $n$. Furthermore, the steepest descent contours for even $n$ lie parallel to the real axis and for odd they lie parallel to the imaginary axis (for $z > 0$). Thus, for the integration contour in \eqref{Boundary-bulk propagator 3}, we pick up the contributions around $t = 0$ and $t = 2\pi$. The two contributions are equal due to symmetry about $t = \pi$, and thus we can expand the integrand about $t = 0$ to quadratic order, leaving us with the following Gaussian integral
\begin{gather}
    \label{Boundary-bulk propagator 4}
    G_{0}^{\phi}(w) \approx \frac{\G(\hphi)^{2}}{\G(2\hphi)}\l(\frac{w}{(1 - \sqrt{1 - w})^{2}}\r)^{\hphi}\int_{-\infty}^{\infty}{\frac{dt}{2\pi}}\exp[-\hphi\frac{\sqrt{1 - w}}{(1 - \sqrt{1 - w})^{2}}t^{2}].
\end{gather}
Performing the $t$ integration and using $\frac{w}{1 - \sqrt{1 - w}} = 1 + \sqrt{1 - w}$, we get the following expression
\begin{gather}
    \label{Boundary-bulk propagator 5}
    G_{0}^{\phi}(w) \approx \frac{1}{\sqrt{4\pi\hphi}}\frac{\G(\hphi)^{2}}{\G(2\hphi)}\sqrt{\frac{(1 - \sqrt{1 - w})^{2}}{\sqrt{1 - w}}}\exp[\hphi\log(\frac{1 + \sqrt{1 - w}}{1 - \sqrt{1 - w}})].
\end{gather}

The relation between $\hphi$ and the mass of the scalar is $\hphi = \frac{1}{2} + \sqrt{\frac{1}{4} + \frac{\mphi^{2}}{d^{2}}}$, which for large $\mphi$ simplifies to $\hphi = \frac{\mphi}{d} + \frac{1}{2} + O(\frac{1}{\mphi})$. Taking a large $\mphi$ limit of \eqref{Boundary-bulk propagator 5} yields
\begin{gather}
    \label{Boundary-bulk propagator 6}
    G_{0}^{\phi}(w) \sim 4^{-\frac{\mphi}{d}}\sqrt{\frac{w}{\sqrt{1 - w}}}\exp[\frac{\mphi}{d}\log(\frac{1 + \sqrt{1 - w}}{1 - \sqrt{1 - w}})],
\end{gather}
where we have suppressed $\frac{1}{\mphi}$ corrections, overall numerical factors, and powers of $\mphi$. Since $\ell_{h} = \frac{2}{d}\log2$ and $\tanh^{-1}\sqrt{1 - w} = \frac{1}{2}\log(\frac{1 + \sqrt{1 - w}}{1 - \sqrt{1 - w}})$, we get the geodesic distance in the exponent of the propagator, thus reproducing the first two terms in the WKB expansion of the boundary-bulk propagator
\begin{gather}
    \label{Boundary-bulk propagator 7}
    G_{0}^{\phi}(w) \sim \sqrt{\frac{w}{\sqrt{1 - w}}}e^{-\mphi\ell_{w}}.
\end{gather}

% ---------------------------------------------------------------------

{\color{black}
\subsection{BTZ propagators at finite momentum}
The three dimensional BTZ black hole has been studied extensively in the context of holography. A special feature of this background is that the Klein-Gordon equation can be solved exactly in terms of an hypergeometric function even at non-vanishing momentum. We use exact expressions of the boundary-bulk and bulk-bulk propagators and show that in the large mass limit they agree with the leading order WKB forms given in \eqref{Boundary-bulk propagator 1} and \eqref{Bulk-bulk propagator 1}. \\

Euclidean BTZ black hole is defined by the metric \eqref{Black brane metric 1} for $d = 2$. We can simultaneously study the BTZ black brane, for which the coordinate $x \in \bbr$, and the BTZ-Schwarzschild black hole, which is characterized by a further periodic identification $x \sim x + 2\pi$. The only difference is that in the latter case the transverse momentum is quantized. Our momentum space analysis does not differentiate between the two cases. \\

We begin by writing the Klein-Gordon equation in Fourier space as in \eqref{KG equation 2} with $d = 2$. The Euclidean boundary-bulk propagator is defined as the solution to this equation that is regular at the horizon. The exact form up to an overall normalization is
\begin{gather}
    \label{BTZ boundary propagator 1}
    G_{k}^{\phi}(z) \sim z^{\Dphi}\l(1 - \frac{z^{2}}{z_{h}^{2}}\r)^{\frac{\abs{\o}z_{h}}{2}}\hf\l(a,b;c;1 - \frac{z^{2}}{z_{h}^{2}}\r),
\end{gather}
with
\begin{gather}
    \label{BTZ boundary propagator 2}
    a = \frac{\Dphi}{2} + \frac{z_{h}}{2}(\abs{\o} + ik), \quad b = \frac{\Dphi}{2} + \frac{z_{h}}{2}(\abs{\o} - ik), \quad c = 1 + \abs{\o}z_{h},
\end{gather}
where $\Dphi = 1 + \sqrt{1 + \mphi^{2}}$. The fact that \eqref{BTZ boundary propagator 1} is a solution can be checked by direct substitution. The regularity at the horizon follows from the Matsubara frequency $\o$ being quantized in units of $\frac{1}{z_{h}} = \frac{2\pi}{\beta}$. To see this more explicitly, we change variables to $\sech\rho = \frac{z}{z_{h}}$ so that near the horizon ($\rho = 0$) the propagator goes as $G_{k}^{\phi}(z) \sim \rho^{\abs{\o}z_{h}}$. \\

Apart from the solution regular at the horizon, the Klein-Gordon equation also has a solution regular at the boundary ($z = 0$). This is the normalizable solution. Up to a normalization, it is given in Fourier space as
\begin{gather}
    \label{BTZ normalizable 1}
    g_{k}^{\phi,\rm norm}(z) \sim z^{\Dphi}\l(1 - \frac{z^{2}}{z_{h}^{2}}\r)^{\frac{\abs{\o}z_{h}}{2}}\hf\l(a,b;\Dphi;\frac{z^{2}}{z_{h}^{2}}\r),
\end{gather}
where $a,b$ were given in \eqref{BTZ boundary propagator 2}. Near the boundary, this solution goes as $g_{k}^{\phi,\rm norm}(z) \sim z^{\Dphi}$. \\

The Euclidean bulk-bulk propagator is a Green's function for the Klein-Gordon equation that is regular at the horizon as well as the boundary. It is directly constructed using the two solutions $G_{k}^{\phi}(z)$ and $g_{k}^{\phi,\rm norm}(z)$ as follows
\begin{gather}
    \label{BTZ bulk propagator 1}
    \mcg_{k}^{\phi}(z_{1};z_{2}) \sim G_{k}^{\phi}(z_{1})g_{k}^{\phi,\rm norm}(z_{2})\Theta(z_{1} > z_{2}) + G_{k}^{\phi}(z_{2})g_{k}^{\phi,\rm norm}(z_{1})\Theta(z_{1} < z_{2}).
\end{gather}

Our goal now is to derive a large mass ($\mphi$) limit for the boundary-bulk and bulk-bulk propagators using the exact expressions \eqref{BTZ boundary propagator 1} and \eqref{BTZ bulk propagator 1}. We will show that the limit presents itself as an asymptotic expansion in $\frac{1}{\mphi}$ such that the leading order result matches with the respective WKB expressions \eqref{Boundary-bulk propagator 1} and \eqref{Bulk-bulk propagator 1} obtained directly from the wave equation. In particular, we will demonstrate that in this limit the momentum dependence ($\o,k$) appears only at orders $\frac{1}{\mphi}$ and higher compared to the leading terms. \\

We note from \eqref{BTZ boundary propagator 1} and \eqref{BTZ normalizable 1} that a large mass limit corresponds to expanding the hypergeometric function for large parameter values. Such expansions have been studied before. Here, we only give the necessary steps required to obtain the asymptotic expansion for the hypergeometric. Interested readers can check the relevant references for detailed proofs and numerical analysis of the formulas derived. \\

We start by approximating the hypergeometric function in boundary-bulk propagator \eqref{BTZ boundary propagator 1}. From \eqref{BTZ boundary propagator 2}, we see that in the limit of large mass the parameters $a$ and $b$ are large. The corresponding expansion was studied in section 2.3 of \cite{Cvitkovic:2016} (also see section 4 of \cite{Paris:2013first}). The idea is to use the following integral representation for the hypergeometric (adapted from equation 15.6.2 in \cite{NIST:DLMF})
\begin{gather}
    \label{HF analysis 1}
    \hf(a,b;c;y) \sim \int_{\mcc}{dt}\exp[H(t)], \\
    \label{HF analysis 2}
    H(t) = (b - 1)\log t + (c - b - 1)\log(t - 1) - a\log(1 - yt).
\end{gather}
Overall factors have been suppressed. For our problem, the parameters $a,b,c$ are as defined in \eqref{BTZ boundary propagator 2} and $y = 1 - \frac{z^{2}}{z_{h}^{2}}$. The contour of integration starts at $t = 0$, goes counter-clockwise around the branch point at $t = 1$ while avoiding the point $t = \frac{1}{y}$ and loops back to $t = 0$ (see Figure 2.6 in \cite{Cvitkovic:2016}). The exponent \eqref{HF analysis 2} has two saddle points
\begin{gather}
    \label{HF analysis 3}
    t_{\pm} = \frac{y(a - b + 1) + 2 - c \pm \sqrt{(y(a - b + 1) + 2 - c)^{2} + 4y(b - 1)(a + 2 - c)}}{2y(a + 2 - c)}.
\end{gather}
In the limit $\mphi \to \infty$ we have $a \sim b \sim \frac{\mphi}{2}$ so that the saddles are approximately located at
\begin{gather}
    \label{HF analysis 4}
    t_{\pm} \to \pm\frac{1}{\sqrt{y}}.
\end{gather}
The value of the exponent at the two saddles goes as
\begin{gather}
    \label{HF analysis 5}
    H(t_{\pm}) = -\mphi\log(1 \mp \sqrt{y}) + \cdots
\end{gather}
so that $t_{+}$ is the dominant saddle (see \cite{Cvitkovic:2016} for details on the contour deformation). The Gaussian integration around the saddle computes the leading correction (again suppressing overall numerical factors)
\begin{gather}
    \label{HF analysis 6}
    \hf(a,b;c;y) \sim \exp[H(t_{+}) - \frac{1}{2}\log H''(t_{+}) + \cdots].
\end{gather}
A direct evaluation of the right hand side with the parameters defined in \eqref{BTZ boundary propagator 2} yields the following result in the large $\mphi$ limit
\begin{gather}
    \label{HF analysis 7}
    \hf(a,b;c;y) \sim \exp[-\mphi\log(1 - \sqrt{y}) - \frac{\abs{\o}z_{h}}{2}\log y - \frac{1}{4}\log y + O\l(\frac{1}{\mphi}\r)].
\end{gather}
For a numerical analysis of the agreement between the steepest descent result and full hypergeometric, see Figure 7 of \cite{Cvitkovic:2016} as well as section 4.4 of \cite{Paris:2013first}. \\

We now put \eqref{HF analysis 7} back into the expression for the propagator \eqref{BTZ boundary propagator 1} while using $y = 1 - \frac{z^{2}}{z_{h}^{2}}$. Notice that the $\abs{\o}$ terms cancel off, thus affirming the fact that in the WKB limit frequency dependence appears at a subleading order. Further, identifying $w = \frac{z^{2}}{z_{h}^{2}}$, we write
\begin{gather}
    \label{BTZ boundary propagator 3}
    G_{k}^{\phi}(z) \sim \exp[\frac{\mphi}{2}\log(\frac{1 + \sqrt{1 - w}}{1 - \sqrt{1 - w}}) + \frac{1}{4}\log(\frac{w^{2}}{1 - w})] = \sqrt{\frac{w}{\sqrt{1 - w}}}e^{-\mphi\ell_{w}},
\end{gather}
where in the last equality we have used the identity $\tanh^{-1}x = \frac{1}{2}\log(\frac{1 + x}{1 - x})$ and the expression \eqref{Regularized length 1} for the geodesic distance,. Hence, we have obtained the desired match with WKB form \eqref{Boundary-bulk propagator 1}. \\

Next, we have to approximate the hypergeometric function in the normalizable solution \eqref{BTZ normalizable 1}. In the limit of large mass, all parameters of the hypergeometric are large. The relevant asymptotic expansion for the hypergeometric in this case was obtained in section 2 of \cite{Paris:2013ugv}. The approach is the same as before. Suppressing overall factors, we use the following integral representation for the hypergeometric (equation 15.6.1 in \cite{NIST:DLMF})
\begin{gather}
    \label{HF analysis 8}
    \hf(a,b;c;y) \sim \int_{0}^{1}{dt}\exp[I(t)], \\
    \label{HF analysis 9}
    I(t) = (b - 1)\log t + (c - b - 1)\log(1 - t) - a\log(1 - yt).
\end{gather}
For our purpose, $y = \frac{z^{2}}{z_{h}^{2}}$ and the parameters $a,b$ are given by \eqref{BTZ boundary propagator 2} while $c = \Dphi$. Note the similarity between the above integrand and the one in \eqref{HF analysis 1}-\eqref{HF analysis 2}. In particular, the exponents $I(t)$ and $H(t)$ differ only by a constant. As a consequence, the saddle points of $I(t)$ are also given by \eqref{HF analysis 3}. For $2a \sim 2b \sim c \sim \mphi$ large, the saddle points are approximately located at
\begin{gather}
    \label{HF analysis 10}
    t_{\pm} \to \frac{1}{1 \pm \sqrt{1 - y}}
\end{gather}
The exponent at the saddles evaluate to
\begin{gather}
    \label{HF analysis 11}
    I(t_{\pm}) = -\mphi\log(1 \pm \sqrt{1 - y}) + \cdots
\end{gather}
Even though the exponent $I(t_{-})$ is larger, the saddle $t_{+}$ already lies on the integration contour if $0 < y < 1$. Further, for this range of $y$ it can be checked that $I''(t_{+}) < 0$ so that the defining contour in \eqref{HF analysis 8} is already the steepest descent path for the saddle. This means that the hypergeometric in question can be approximated by the Gaussian integration around the $t_{+}$ saddle (see \cite{Paris:2013first} for more details). The result is given as (suppressing overall numerical factors)
\begin{gather}
    \label{HF analysis 12}
    \hf(a,b;c;y) \sim \exp[-\mphi\log(1 + \sqrt{1 - y}) - \frac{\abs{\o}z_{h}}{2}\log(1 - y) - \frac{1}{4}\log(1 - y) + O\l(\frac{1}{\mphi}\r)]
\end{gather}

Numerical checks for formulas similar to \eqref{HF analysis 12} were performed in \cite{Paris:2013first}. We now combine the above expression with the remaining terms in \eqref{BTZ normalizable 1} while recalling that $y = \frac{z^{2}}{z_{h}^{2}}$. We note that for the normalizable solution as well the $\o$ dependence cancels off at leading order. Identifying $w = \frac{z^{2}}{z_{h}^{2}}$ and using $(1 + \sqrt{1 - w})(1 - \sqrt{1 - w}) = w$, we express the solution as
\begin{gather}
    \label{BTZ normalizable 2}
    g_{k}^{\phi,\rm norm}(z) \sim \exp[-\frac{\mphi}{2}\log(\frac{1 + \sqrt{1 - w}}{1 - \sqrt{1 - w}}) + \frac{1}{4}\log(\frac{w^{2}}{1 - w})] = \sqrt{\frac{w}{\sqrt{1 - w}}}e^{\mphi\ell_{w}}.
\end{gather}
In the last equality we have used the identity $\tanh^{-1}x = \frac{1}{2}\log(\frac{1 + x}{1 - x})$ and equation \eqref{Regularized length 1} for the geodesic distance. Substituting the expressions \eqref{BTZ boundary propagator 3} and \eqref{BTZ normalizable 2} into the bulk-bulk propagator \eqref{BTZ bulk propagator 1} reproduces the WKB form derived in \eqref{Bulk-bulk propagator 1}.
}

% ---------------------------------------------------------------------

\section{One point Function}
\label{Section: one point function}
\begin{figure}[!t]
    \centering
    \begin{tikzpicture}[x=1pt,y=1pt,yscale=-1,xscale=1]
    % \path (0,451); %set diagram left start at 0, and has height of 451
    
    % Base boundary line
    \draw [line width = 1.5] (100,110) -- (240,110);
    \draw (80,110) node {$w = 0$};
    
    % Low opacity boundary lines
    \draw [opacity = 0.3, line width = 0.75] (100,110) -- (110,100);
    \draw [opacity = 0.3, line width = 0.75] (120,110) -- (130,100);
    \draw [opacity = 0.3, line width = 0.75] (140,110) -- (150,100);
    \draw [opacity = 0.3, line width = 0.75] (160,110) -- (170,100);
    \draw [opacity = 0.3, line width = 0.75] (180,110) -- (190,100);
    \draw [opacity = 0.3, line width = 0.75] (200,110) -- (210,100);
    \draw [opacity = 0.3, line width = 0.75] (220,110) -- (230,100);
    \draw [opacity = 0.3, line width = 0.75] (240,110) -- (250,100);
    
    % Heavy scalar boundary-bulk propagator
    \draw [line width = 1] (170,110) -- (170,169);
    \draw (160,140) node {$G_{0}^{\phi}$};
    
    % Heavy scalar-Weyl squared interaction vertex
    \draw (170,170) node [font = \LARGE, color={rgb, 255:red, 0; green, 200; blue, 255}] {$\bm{\times}$};
    \draw (170,182) node [font = \footnotesize] {$W^{2}(w)$};
    \end{tikzpicture}
    
    \caption{Witten diagram for thermal CFT one point function of a primary being sourced holographically by a coupling to Weyl squared term. The interaction vertex has been marked and the bulk point is integrated over the entire exterior region. Boundary-bulk propagator of the dual scalar is denoted by a solid line.}
    \label{Witten diagram 5}
\end{figure}
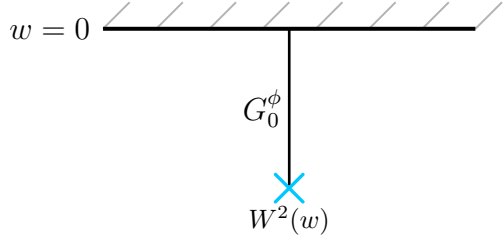

In this appendix, we use the notation and expressions developed in section \ref{Section: preliminaries} for reviewing the calculation of the thermal one point function \eqref{One point function 1}, first performed in \cite{Grinberg:2020fdj}. We consider one point function being sourced by the coupling of the bulk heavy scalar to Weyl tensor squared, as in \eqref{Action 5}. The Witten diagram for this contribution is shown in Figure \ref{Witten diagram 5}. The computation at large values of $\mphi$, or equivalently $\hphi$, can be performed in two ways --- by exactly performing the integral and then taking a limit, or by using the WKB expression and evaluating the integral using saddle point method. The match between the two engenders support for the use of WKB and saddle point method for bulk integrals. \\

The contribution of the Witten diagram reads as follows
\begin{gather}
    \label{One point function 2}
    \ev{\ophi(0)}_{\beta} = \alpha\int_{0}^{1}{\frac{dw}{w^{2}}}G_{0}^{\phi}(w)W^{2}(w),
\end{gather}
which reduces to the following integral after substituting the value \eqref{Weyl squared 1} for the Weyl tensor squared.
\begin{gather}
    \label{One point function 3}
    \ev{\ophi(0)}_{\beta} \sim \alpha\int_{0}^{1}{dw}G_{0}^{\phi}(w).
\end{gather}
The one point function is evaluated at vanishing momentum because of the translational isometries.

\subsection{Exact calculation}
Since the propagator in \eqref{One point function 2} is at vanishing momentum, we can use the exact form given in \eqref{Boundary-bulk propagator 2}. Substituting this yields the following integral
\begin{gather}
    \label{One point function 4}
        \ev{\ophi(0)}_{\beta} \sim \alpha\frac{\G(\hphi)^{2}}{\G(2\hphi)}\int_{0}^{1}{dw}\,w^{\hphi}\hf(\hphi,\hphi;1;1 - w).
\end{gather}
This integral is divergent for $\Re(\hphi) > 2$ but can be performed exactly otherwise. It is an example of the integral formula in equation 7.511-4 in \cite{tableofintegrals}, which reads
\begin{gather}
    \label{Integral formula 1}
    \int_{0}^{1}{dx}x^{\g - 1}(1 - x)^{\rho - 1}\hf(\alpha,\beta;\g;x) = \frac{\G(\g)\G(\rho)\G(\g + \rho - \alpha - \beta)}{\G(\g + \rho - \alpha)\G(\g + \rho - \beta)}.
\end{gather}
Taking the integration variable as $x = 1 - w$ and other parameters to be $\alpha = \beta = \hphi$, $\g = 1$, and $\rho = \hphi + 1$ produces the integral in \eqref{One point function 4}. Thus, we can take the result of \eqref{Integral formula 1} for $\Re(\hphi) < 2$ and analytically continue the answer for large $\hphi$ \cite{Freedman:1998tz}. Then, the result of the integration is given as
\begin{gather}
    \label{Integral formula 2}
    \int_{0}^{1}{dw}\,w^{\hphi}\hf(\hphi,\hphi;1;1 - w) = \G(\hphi + 1)\G(2 - \hphi) = \hphi(1 - \hphi)\frac{\pi}{\sin\pi\hphi},
\end{gather}
where in the last equality we have made use of the identities $\G(x + 1) = x\G(x)$ and $\G(x)\G(1 - x) = \frac{\pi}{\sin\pi x}$. \\

The one point function is now given as
\begin{gather}
    \label{One point function 5}
    \ev{\ophi(0)}_{\beta} \sim \alpha\frac{\G(\hphi)^{2}}{\G(2\hphi)}\frac{\pi\hphi(1 - \hphi)}{\sin\pi\hphi}.
\end{gather}
We see that the one point function is a well defined meromorphic function of the conformal dimension of the primary (and hence the mass of the bulk field). The poles at $\hphi = n$ ($\Dphi = nd$) for integers $n > 2$ correspond to mixing of $\ophi$ with powers of the CFT stress tensor $T^{n}$ (see \cite{Berenstein:2022nlj} and \cite{Grinberg:2020fdj} for further discussion). \\

We want to study the large $\hphi$, or equivalently, large $\mphi$ limit of the one point function. The Gamma functions, at leading order, produce the factor of horizon distance in the exponent
\begin{gather}
    \label{One point function 6}
    \frac{\G(\hphi)^{2}}{\G(2\hphi)} \sim 2^{-2\hphi} \sim e^{-\mphi\ell_{h}}.
\end{gather}
To extract a limit out of the $\sin$ factor in the denominator of \eqref{One point function 5}, a negative imaginary part is given to $\hphi$. At leading order, we can approximate $\frac{1}{\sin\pi\hphi} \sim 2e^{-i\pi\hphi} \sim 2e^{-i\mphi\tau_{s}}$, where we have identified the factor of proper radial time to singularity. Combining this with \eqref{One point function 6} reproduces the exponential behaviour claimed earlier
\begin{gather}
    \label{One point function 7}
    \ev{\ophi(0)}_{\beta} \sim \alpha\exp[-\mphi(\ell_{h} + i\tau_{s})].
\end{gather}

\subsection{Saddle point method}
Our aim here is to reproduce the form \eqref{One point function 7} for the thermal one point function using the WKB expression for the boundary-bulk propagator at vanishing momentum and then approximating the bulk integral using saddle point method. Using the WKB form of the propagator in \eqref{Boundary-bulk propagator 1}, the integral \eqref{One point function 3} for the thermal one point function is written as
\begin{gather}
    \label{One point function 8}
    \ev{\ophi}_{\beta} \sim \alpha\int_{0}^{1}{dw}\exp[-\mphi\ell_{w} + \frac{1}{4}\log(\frac{w^{2}}{1 - w})].
\end{gather}
The saddle equation for the integrand reads
\begin{gather}
    \label{One point saddle 1}
    -\frac{\mphi}{d}\frac{1}{w^{*}\sqrt{1 - w^{*}}} + \frac{2 - w^{*}}{4w^{*}(1 - w^{*})} = 0,
\end{gather}
which is solved at large $\mphi$ to give
\begin{gather}
    \label{One point saddle 2}
    w^{*} \approx -\l(\frac{4\mphi}{d}\r)^{2}.
\end{gather}
As $\mphi$ is taken the be large, the saddle lies at $\infty$. Giving $\mphi$ a negative imaginary part defines the proper length at the saddle as
\begin{gather}
    \label{Regularized length 4}
    \ell_{w^{*}} \approx \ell_{h} + i\tau_{s}.
\end{gather}
So, at leading order, the exponential contribution to the one point function is read off from the saddle point answer as
\begin{gather}
    \label{One point function 9}
    \ev{\ophi}_{\beta} \sim \alpha\exp[-\mphi\ell_{w^{*}}] \sim \alpha\exp[-\mphi(\ell_{h} + i\tau_{s})].
\end{gather}
It matches exactly with the form obtained from the exact computation of the bulk integral. A steepest descent analysis for the one-point function was provided in \cite{Grinberg:2020fdj} and follows the approach of section \ref{Section: SDA 1}.

\bibliographystyle{JHEP}
\bibliography{References}

@article{Grinberg:2020fdj,
    author = "Grinberg, Matan and Maldacena, Juan",
    title = "{Proper time to the black hole singularity from thermal one-point functions}",
    eprint = "2011.01004",
    archivePrefix = "arXiv",
    primaryClass = "hep-th",
    doi = "10.1007/JHEP03(2021)131",
    journal = "JHEP",
    volume = "03",
    pages = "131",
    year = "2021"
}

@article{David:2022nfn,
    author = "David, Justin R. and Kumar, Srijan",
    title = "{Thermal one point functions, large d and interior geometry of black holes}",
    eprint = "2212.07758",
    archivePrefix = "arXiv",
    primaryClass = "hep-th",
    doi = "10.1007/JHEP03(2023)256",
    journal = "JHEP",
    volume = "03",
    pages = "256",
    year = "2023"
}

@article{David:2023uya,
    author = "David, Justin R. and Kumar, Srijan",
    title = "{Thermal one-point functions: CFT\textquoteright{}s with fermions, large d and large spin}",
    eprint = "2307.14847",
    archivePrefix = "arXiv",
    primaryClass = "hep-th",
    doi = "10.1007/JHEP10(2023)143",
    journal = "JHEP",
    volume = "10",
    pages = "143",
    year = "2023"
}

@article{Louko_2000,
   title={Geodesic propagators and black hole holography},
   volume={62},
   ISSN={1089-4918},
   url={http://dx.doi.org/10.1103/PhysRevD.62.044041},
   DOI={10.1103/physrevd.62.044041},
   number={4},
   journal={Physical Review D},
   publisher={American Physical Society (APS)},
   author={Louko, Jorma and Marolf, Donald and Ross, Simon F.},
   year={2000}}

@article{Kraus:2002iv,
    author = "Kraus, Per and Ooguri, Hirosi and Shenker, Stephen",
    title = "{Inside the horizon with AdS / CFT}",
    eprint = "hep-th/0212277",
    archivePrefix = "arXiv",
    reportNumber = "UCLA-02-TEP-41, CALT-68-2421, SU-ITP-02-45",
    doi = "10.1103/PhysRevD.67.124022",
    journal = "Phys. Rev. D",
    volume = "67",
    pages = "124022",
    year = "2003"
}

@article{Fidkowski:2003nf,
    author = "Fidkowski, Lukasz and Hubeny, Veronika and Kleban, Matthew and Shenker, Stephen",
    title = "{The Black hole singularity in AdS / CFT}",
    eprint = "hep-th/0306170",
    archivePrefix = "arXiv",
    reportNumber = "SU-ITP-03-16",
    doi = "10.1088/1126-6708/2004/02/014",
    journal = "JHEP",
    volume = "02",
    pages = "014",
    year = "2004"
}

@article{Festuccia:2005pi,
    author = "Festuccia, Guido and Liu, Hong",
    title = "{Excursions beyond the horizon: Black hole singularities in Yang-Mills theories. I.}",
    eprint = "hep-th/0506202",
    archivePrefix = "arXiv",
    reportNumber = "MIT-CTP-3641",
    doi = "10.1088/1126-6708/2006/04/044",
    journal = "JHEP",
    volume = "04",
    pages = "044",
    year = "2006"
}

@article{Rodriguez-Gomez:2021pfh,
    author = "Rodriguez-Gomez, D. and Russo, J. G.",
    title = "{Correlation functions in finite temperature CFT and black hole singularities}",
    eprint = "2102.11891",
    archivePrefix = "arXiv",
    primaryClass = "hep-th",
    reportNumber = "ICCUB-21-002",
    doi = "10.1007/JHEP06(2021)048",
    journal = "JHEP",
    volume = "06",
    pages = "048",
    year = "2021"
}

@article{Rodriguez-Gomez:2021mkk,
    author = "Rodriguez-Gomez, D. and Russo, J. G.",
    title = "{Thermal correlation functions in CFT and factorization}",
    eprint = "2105.13909",
    archivePrefix = "arXiv",
    primaryClass = "hep-th",
    reportNumber = "ICCUB-21-007",
    doi = "10.1007/JHEP11(2021)049",
    journal = "JHEP",
    volume = "11",
    pages = "049",
    year = "2021"
}

@article{Georgiou:2022ekc,
    author = "Georgiou, George and Zoakos, Dimitrios",
    title = "{Holographic correlation functions at finite density and/or finite temperature}",
    eprint = "2209.14661",
    archivePrefix = "arXiv",
    primaryClass = "hep-th",
    doi = "10.1007/JHEP11(2022)087",
    journal = "JHEP",
    volume = "11",
    pages = "087",
    year = "2022"
}

@article{Georgiou:2023xpg,
    author = "Georgiou, George and Zoakos, Dimitrios",
    title = "{Holographic three-point correlators at finite density and temperature}",
    eprint = "2309.07645",
    archivePrefix = "arXiv",
    primaryClass = "hep-th",
    doi = "10.1007/JHEP12(2023)125",
    journal = "JHEP",
    volume = "12",
    pages = "125",
    year = "2023"
}

@article{Horowitz:2023ury,
    author = "Horowitz, Gary T. and Leung, Henry and Queimada, Leonel and Zhao, Ying",
    title = "{Boundary signature of singularity in the presence of a shock wave}",
    eprint = "2310.03076",
    archivePrefix = "arXiv",
    primaryClass = "hep-th",
    journal = "",
    year = "2023"
}

@article{Berenstein:2022nlj,
    author = "Berenstein, David and Mancilla, Robinson",
    title = "{Aspects of thermal one-point functions and response functions in AdS black holes}",
    eprint = "2211.05144",
    archivePrefix = "arXiv",
    primaryClass = "hep-th",
    doi = "10.1103/PhysRevD.107.126010",
    journal = "Phys. Rev. D",
    volume = "107",
    number = "12",
    pages = "126010",
    year = "2023"
}

@article{Freedman:1998tz,
    author = "Freedman, Daniel Z. and Mathur, Samir D. and Matusis, Alec and Rastelli, Leonardo",
    title = "{Correlation functions in the CFT(d) / AdS(d+1) correspondence}",
    eprint = "hep-th/9804058",
    archivePrefix = "arXiv",
    reportNumber = "MIT-CTP-2727",
    doi = "10.1016/S0550-3213(99)00053-X",
    journal = "Nucl. Phys. B",
    volume = "546",
    pages = "96--118",
    year = "1999"
}

@article{Ceplak:2024bja,
    author = "\v{C}eplak, Nejc and Liu, Hong and Parnachev, Andrei and Valach, Samuel",
    title = "{Black Hole Singularity from OPE}",
    eprint = "2404.17286",
    archivePrefix = "arXiv",
    journal = "",
    primaryClass = "hep-th",
    year = "2024"
}

@article{Krishna:2021fus,
    author = "Krishna, Hare and Rodriguez-Gomez, D.",
    title = "{Holographic thermal correlators revisited}",
    eprint = "2108.00277",
    archivePrefix = "arXiv",
    primaryClass = "hep-th",
    doi = "10.1007/JHEP11(2021)139",
    journal = "JHEP",
    volume = "11",
    pages = "139",
    year = "2021"
}

@article{Amado:2008hw,
    author = "Amado, Irene and Hoyos-Badajoz, Carlos",
    title = "{AdS black holes as reflecting cavities}",
    eprint = "0807.2337",
    archivePrefix = "arXiv",
    primaryClass = "hep-th",
    reportNumber = "IFT-UAM-CSIC-08-45",
    doi = "10.1088/1126-6708/2008/09/118",
    journal = "JHEP",
    volume = "09",
    pages = "118",
    year = "2008"
}

@article{Jana:2020vyx,
    author = "Jana, Chandan and Loganayagam, R. and Rangamani, Mukund",
    title = "{Open quantum systems and Schwinger-Keldysh holograms}",
    eprint = "2004.02888",
    archivePrefix = "arXiv",
    primaryClass = "hep-th",
    doi = "10.1007/JHEP07(2020)242",
    journal = "JHEP",
    volume = "07",
    pages = "242",
    year = "2020"
}

@book{tableofintegrals,
    author = "Gradshteyn, Izrail Solomonovich and Ryszhik, Iosif Moiseevich",
    title = "{Table of integrals, series, and products}",
    publisher = "Elsevier/Academic Press, Amsterdam, seventh ed., 2007. Translated from the Russian,
    Translation edited and with a preface by Alan Jeffrey and Daniel Zwillinger, With one
    CD-ROM (Windows, Macintosh and UNIX)."
}

@inbook{Freedman_Van_Proeyen_2012,
    place={Cambridge},
    title={The {AdS/CFT} correspondence},
    booktitle={Supergravity},
    publisher={Cambridge University Press},
    author={Freedman, Daniel Z. and Van Proeyen, Antoine},
    year={2012},
    pages={527–572}
}

@book{kawai2005algebraic,
  title={Algebraic Analysis of Singular Perturbation Theory},
  author={Kawai, T. and Takei, Y.},
  isbn={9780821835470},
  lccn={2005048161},
  series={Iwanami series in modern mathematics},
  url={https://books.google.co.in/books?id=GDsjJLinxogC},
  year={2005},
  publisher={American Mathematical Society}
}

@ARTICLE{Cvitkovic:2016,
       author = {{Cvitkovi{\'c}}, Mislav and {Smith}, Ana-Sun{\v{c}}ana and {Pande}, Jayant},
        title = "{Asymptotic expansions of the hypergeometric function with two large parameters{\textemdash}application to the partition function of a lattice gas in a field of traps}",
      journal = {Journal of Physics A Mathematical General},
         year = 2017,
        month = jun,
       volume = {50},
       number = {26},
          eid = {265206},
        pages = {265206},
          doi = {10.1088/1751-8121/aa7213},
archivePrefix = {arXiv},
       eprint = {1602.05146},
 primaryClass = {math-ph},
       adsurl = {https://ui.adsabs.harvard.edu/abs/2017JPhA...50z5206C}
}

@misc{NIST:DLMF,
         key = "{\relax DLMF}",
       title = "{\it NIST Digital Library of Mathematical Functions}",
howpublished = "\url{https://dlmf.nist.gov/}, Release 1.2.4 of 2025-03-15",
         url = "https://dlmf.nist.gov/",
        note = "F.~W.~J. Olver, A.~B. {Olde Daalhuis}, D.~W. Lozier, B.~I. Schneider,
                R.~F. Boisvert, C.~W. Clark, B.~R. Miller, B.~V. Saunders,
                H.~S. Cohl, and M.~A. McClain, eds."}

@article{Birmingham:1998nr,
    author = "Birmingham, Danny",
    title = "{Topological black holes in Anti-de Sitter space}",
    eprint = "hep-th/9808032",
    archivePrefix = "arXiv",
    doi = "10.1088/0264-9381/16/4/009",
    journal = "Class. Quant. Grav.",
    volume = "16",
    pages = "1197--1205",
    year = "1999"
}

@article{Camporesi:1994ga,
    author = "Camporesi, R. and Higuchi, A.",
    title = "{Spectral functions and zeta functions in hyperbolic spaces}",
    doi = "10.1063/1.530850",
    journal = "J. Math. Phys.",
    volume = "35",
    pages = "4217--4246",
    year = "1994"
}

@article{Paris:2013first,
    author = "Paris, R. B.",
    title = "{Asymptotics of the Gauss hypergeometric function with large parameters, I}",
    year = "2013",
    doi = "10.7153/jca-02-15",
    volume = "2",
    pages = "183--203",
    journal = "Journal of Classical Analysis",
    number = "2",
}

@article{Paris:2013ugv,
    author = "Paris, R. B.",
    title = "{Asymptotics of the Gauss hypergeometric function with large parameters, II}",
    doi = "10.7153/jca-03-01",
    journal = "J. Class. Anal.",
    volume = "3",
    number = "1",
    pages = "1--15",
    year = "2013"
}

\end{document}